\documentclass[aps,prx,amsmath,amssymb,preprint]{revtex4-1}

\usepackage{epsfig}
\usepackage{graphicx}
\usepackage{amsmath}
\usepackage{color}
\usepackage{float}
\makeatletter
\@ifundefined{textcolor}{}
{%
 \definecolor{BLACK}{gray}{0}
 \definecolor{WHITE}{gray}{1}
 \definecolor{RED}{rgb}{1,0,0}
 \definecolor{GREEN}{rgb}{0,1,0}
 \definecolor{BLUE}{rgb}{0,0,1}
 \definecolor{CYAN}{cmyk}{1,0,0,0}
 \definecolor{MAGENTA}{cmyk}{0,1,0,0}
 \definecolor{YELLOW}{cmyk}{0,0,1,0}
}

\usepackage{graphics}\usepackage{subfig}
\newcommand{\eps}{\varepsilon}

\def\vE{\mathbf{E}}

\def\vM{\mathbf{M}}
\def\vN{\mathbf{N}}
\def\e{\begin{equation}}
\def\f{\end{equation}}
\def\_#1{{\bf #1}}

\def\E{\varepsilon}
\def\M{\mu}

\def\.{\cdot}

\def\Re{{\rm Re\mit}}

\makeatother

\begin{document}

\title{Tunable scattering cancellation cloak with plasmonic ellipsoids in the visible}

\author{Martin Fruhnert\footnote{martin.fruhnert@kit.edu}}
\affiliation{Institute of Theoretical Solid State Physics, Karlsruhe Institute of Technology, Wolfgang-Gaede-Strasse 1, 76131 Karlsruhe, Germany}

\author{Alessio Monti}
\affiliation{Niccol\`o Cusano University, Via Don Carlo Gnocchi 3, Rome 00166, Italy}

\author{Ivan Fernandez-Corbaton}
\affiliation{Institute of Nanotechnology, Karlsruhe Institute of Technology, P.O. Box 3640, 76021 Karlsruhe, Germany}

\author{Andrea Al\`u}
\affiliation{Department of Electrical and Computer Engineering, The University of Texas at Austin, Austin, TX 78712, USA}

\author{Alessandro Toscano}
\affiliation{Department of Engineering, Roma Tre University, Via Vito Volterra 62, Rome 00154, Italy}

\author{Filiberto Bilotti}
\affiliation{Department of Engineering, Roma Tre University, Via Vito Volterra 62, Rome 00154, Italy}

\author{Carsten Rockstuhl}
\affiliation{Institute of Theoretical Solid State Physics, Karlsruhe Institute of Technology, Wolfgang-Gaede-Strasse 1, 76131 Karlsruhe, Germany}
\affiliation{Institute of Nanotechnology, Karlsruhe Institute of Technology, P.O. Box 3640, 76021 Karlsruhe, Germany}



\begin{abstract}
The scattering cancellation technique is a powerful tool to reduce the scattered field from electrically small objects in a specific frequency window. The technique relies on covering the object of interest with a shell that scatters light into the far field of equal strength as the object, but $\pi$ out-of-phase. The resulting destructive interference prohibits its detection in measurements that probe the scattered light. Whereas at radio or microwave frequencies feasible designs have been proposed that allow to tune the operational frequency upon request, similar capabilities have not yet been explored in the visible. However, such ability is decisive to capitalize on the technique in many envisioned applications. Here, we solve the problem and study the use of small metallic nanoparticles with an ellipsoidal shape as the material from which the shell is made to build an isotropic geometry. Changing the aspect ratio of the ellipsoids allows to change the operational frequency. The basic functionality is explored with two complementary analytical approaches. Additionally, we present a powerful multiscattering algorithm that can be used to perform full wave simulations of clusters of arbitrary particles. We utilize this method to analyze the scattering of the presented designs numerically. Hereby we provide useful guidelines for the fabrication of this cloak with self-assembly methods by investigating the effects of disorder.
\end{abstract}
\maketitle

\section{Introduction}
With today's possibilities in designing and fabricating complicated nanostructures from different materials, entirely new approaches to tame the propagation of light have been pioneered. This brought many new and interesting applications within reach \cite{Yannopapas_transparency,Nordlander_nanoantennas,Wegener_helix,Pawlak,Pawlak2}. Hiding objects from detection with electromagnetic radiation is potentially one of the most fascinating aspects of modern nanooptics. The possibility to conceal an object at a specific frequency was simultaneously proposed by Ulf Leonhardt and Sir John Pendry and co-workers. They suggested to explore different techniques for a coordinate transformation to design a supporting structure, made from materials with complicated properties, to guide an incident electromagnetic field around a predefined spatial region \cite{pendry_cloak,Leonhardt_cloak,Jiang_cloak}. With this approach, it is possible to hide an arbitrary object from an observer if placed inside the transformed region. Applications exist e.g. when cloaking contact fingers and bus bars in solar cells \cite{Schumann_cloaking_solar_cell}.

If a general coordinate transformation is used, metamaterials with an anisotropic and inhomogeneous permittivity and permeability are required to implement the supporting structure. Since making such materials available obviously constitutes a challenge, an eminent question has been how to lower the constraints on the required functionality to simplify the realization of the necessary material properties. In the slip-stream of such research, for example, carpet cloaks were explored that hide an object that resides on a planar surface, e.g.~a metal mirror \cite{Pendry_carpet,Zentgraf_carpet,Wegener_carpet,park_carpet}.

A completely different strategy to hide small particles from detection led to the development of the scattering cancellation technique \cite{alu_engheta_cloak,Silveirinha_cloak,Bilotti_te_tm,monti_cloak1,monti_cloak2}. There, the scattering response of an electrically small object to a given illumination is reduced by encapsulating it with a supporting shell. The shell is designed such that the illumination induces an electric dipole moment of the same magnitude as in the core object. However, the dipole moment shall oscillate with a phase difference of $\pi$ with respect to the dipole moment of the core object with the effect that the emerging destructive interference will cancel the scattered field \cite{Kerker_invisibility}. This renders the object undetectable at the cloaking frequency in techniques that use the scattered light to detect the presence of an object.

This reduction of the scattering cross section, as it has been explored in the past, is a valuable ability that might find many applications \cite{Rockstuhl_book}. Examples are the suppression of the cross-talk between closely spaced nanoantennas \cite{Onal_antenna_cloak}, the suppression of the perturbation of the field to be detected by a tip in a scanning near-field optical microscope operated in scattering mode \cite{Bilotti_tip_cloak}, or the manipulation of optical forces \cite{Bilotti_force_cloak}. To make these applications come true, it is of paramount importance to be able to tune the operational frequency where the scattering response is reduced at a predefined frequency imposed by the specific application.

For an operational frequency in the radio or microwave part of the spectrum, different metasurfaces have been suggested for the implementation of the scattering cancellation technique \cite{alu_invisibility_prb}. Such mantle cloaks, as they were dubbed, can be made from different motifs and their functionality has been verified in pioneering experiments \cite{Alu_experimental_mantle_cloak}. On the contrary, for an application at optical frequencies, thus far only the feasibility of the concept was demonstrated but not yet the ability to tune the operational frequency. These first experiments employed a shell made from spherical metallic nanoparticles to cover the object from which the scattering response should be suppressed, i.e. usually a dielectric sphere \cite{,Muhlig_prb_cloak,Muhlig_selfassembled_cloak}. This dielectric sphere has to be sufficiently small such that the scattering response is dominated by the electric dipole moment. The restriction to spherical metallic nanoparticles eventually fixes the operational frequency of the cloaking effect.

To mitigate this problem of low tunability, we suggest and study here the use of metallic ellipsoidal nanoparticles as the building blocks from which the shell is made. Changing the aspect ratio changes the resonance frequency of the nanoparticles and with that the operational frequency across an extended spectral domain. Since the shell material is usually fabricated by means of self-assembly techniques, an important question is whether the scattering strength can be reduced for a random arrangement and orientation of the ellipsoids on top of the core object. Then, only a fraction of the ellipsoidal nanoparticles will effectively interact with the external illumination that is considered to have a fixed linear polarization. 

Here, we study the basic design of a scattering cancellation device where the shell is made from metallic ellipsoids by two complementary analytical techniques. The purpose of these techniques is to demonstrate the underlying principle that causes the ability to tune the operational frequency and to provide a glimpse on the possible performance. The eventual functionality of the design is demonstrated by full wave numerical simulations. These simulations rely on a powerful tool in where the scattering response of each individual ellipsoid is expressed in terms of a T-matrix \cite{waterman_t-matrix}. The T-matrix relates the incident field expanded into spherical harmonics with the scattered field. Afterwards, a multiple scattering formalism is used to study the self-consistent optical response. This allows to study the influence of the positional and rotational disorder on the performance of the device.        

\section{Theoretical Considerations}
\label{Theoretical considerations}
In this section, we want to study by two different analytical means a scattering structure that consists of a dielectric sphere (the core object) that is covered with a layer of silver ellipsoids (the supporting shell). In the first analytical approach, the covering shell is modeled as a thin reactive metasurface to which a reactance is assigned based on the polarizability of the ellipsoids. Comparing this reactance to the necessary reactance that suppresses the scattering signal allows to identify the parameter region where the scattering is reduced.

In the second analytical approach, the nanoparticles forming the covering shell are considered as an effectively homogeneous medium with effective material parameters. The effective properties of this medium are derived from basic mixing rules in dipolar approximation. Beyond insights into the spectral position of the operational frequency, the approach is useful to get an educated guess on the performance of the device. 

\subsection{Surface Homogenization}
\begin{figure}[htbp]
\centering \includegraphics[width=0.7\textwidth]{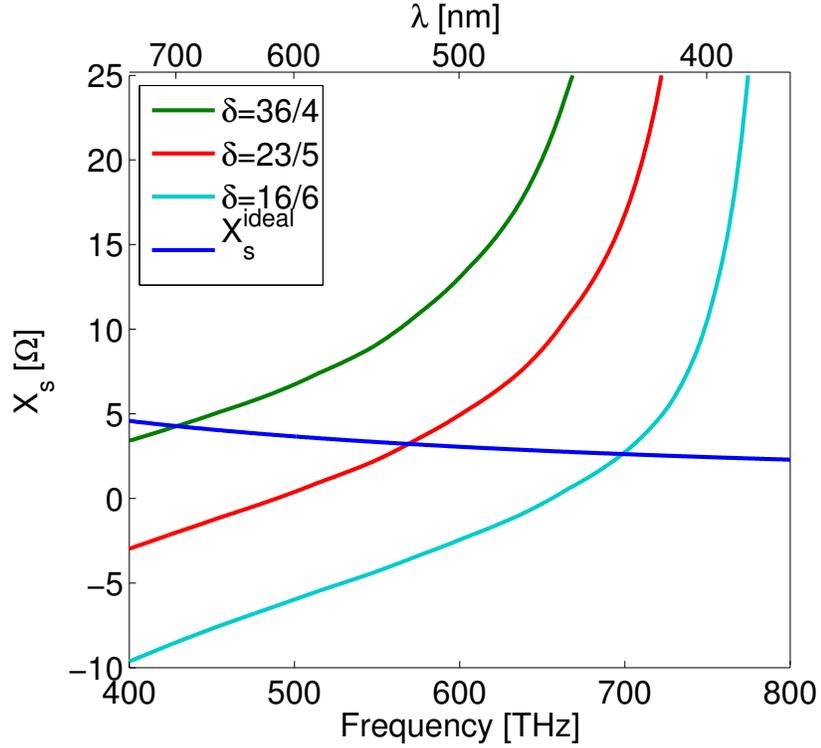} \caption{Surface reactance calculated for different aspect ratios $\delta=a_\textrm{z}/a_\textrm{x}$ of the constituting ellipsoids and the ideal surface reactance that is required to cancel the scattering. Spectral positions where both surface reactances are equal correspond to the operational frequencies for different aspect ratios.}
\label{fig_surface}
\end{figure}
At first, we want to investigate the operating frequency of the proposed scattering cancellation and its tunability. For this purpose, we analyze the dependence of the cloaking frequency on the aspect ratio of the ellipsoids.

We assume that the nanoparticles are electrically very small compared to the wavelength e.g.~$\lambda/a_\textrm{z}\approx20$, where $a_\textrm{z}$ is the major semi axis of the ellipsoid. Therefore, the shell can be modeled as a two dimensional metasurface. The behavior of the shell is then characterized by an average surface reactance. The surface reactance corresponds to the imaginary part of the surface impedance $Z_\textrm{s}(\omega)=R_\textrm{s}(\omega)-iX_\textrm{s}(\omega)$. For an array of particles that are characterized by a polarizability $\alpha$, this surface reactance reads as \cite{saeidi_surface}
\begin{align}
X_\textrm{s}(\omega)=-\frac{d^2}{k}(\Re[1/\alpha(\omega)]-\Re[\beta(\omega)])~.
\label{eq:reactance}
\end{align}
Here, $\beta(\omega)$ describes the interaction of the nanoparticles in the array while being exposed to an external excitation \cite{tretyakov_book}, $d$ defines the distance of the particles, and $k$ is the wave number of the incident plane wave.

The polarizability $\alpha(\omega)$ of an ellipsoid is analytically known to be \cite{bohren}
\begin{align}
\alpha_\textrm{z}(\omega)=\frac{\varepsilon_\textrm{p}(\omega)-\varepsilon_\textrm{h}}{\varepsilon_\textrm{h}+N_\textrm{z}(\varepsilon_\textrm{p}(\omega)-\varepsilon_\textrm{h})}~,
\label{analytical_eiipsoid}
\end{align}
where $\varepsilon_\textrm{p}(\omega)$ is the permittivity of the ellipsoid obtained from experimental data \cite{johnson_christy}, $\varepsilon_\textrm{h}=1$ is the host medium, in our case air, and $N_\textrm{z}$ is the depolarization factor along $z$ which depends on the aspect ratio $\delta=a_\textrm{z}/a_\textrm{x}$ of the ellipsoid. The axes are chosen such that the volume $V=\frac{4}{3}\pi a_\textrm{x}^2  a_\textrm{z}$ stays constant.

When the ellipsoids cover a sphere, they are not in perfect alignment like in a flat array. To account for this disorder, we calculate an effective polarizability by taking the average of the three components $\alpha_\textrm{x}(\omega)$, $\alpha_\textrm{y}(\omega)$, and $\alpha_\textrm{z}(\omega)$. This implies, that the homogeneous medium that is considered here is made from ellipsoids that have an arbitrary orientation in space. The filling fraction is derived from the distance of the particle and the total surface area of the core object. Here, we assume a distance of $d = 40$ nm which reflects a filling fraction of approximately $6.5\%$ that is assumed later on.

The surface reactance that is necessary to suppress the scattering response from a spherical core object is given by \cite{alu_invisibility_prb}
\begin{align}
X_\textrm{s}^\textrm{ideal}(\omega_\textrm{c})=\frac{2[2+\varepsilon_\textrm{sph}-\gamma^3(\varepsilon_\textrm{sph}-1)]}{3\frac{\omega_\textrm{c}}{c} r_\textrm{sph}\gamma^3(\varepsilon_\textrm{sph}-1)}~,
\end{align}
where $\omega_\textrm{c}$ is the desired operational frequency, $\varepsilon_\textrm{sph}=2.1$ is the relative permittivity of the core sphere, $\gamma=r_\textrm{sph}/r_\textrm{clo}$, $r_\textrm{sph}=61$ nm is the core radius and $r_\textrm{clo}=66$ nm is the radius of the mantle cloak.

Now, the surface reactance strongly depends on the polarizability $\alpha(\omega)$ of the particles forming the reactive surface. Therefore, it can be tuned by changing the aspect ratio $\delta=a_\textrm{z}/a_\textrm{x}$ of the ellipsoids as depicted in Fig.~\ref{fig_surface}, where we calculate the surface reactance by substituting Eq.~\ref{analytical_eiipsoid} into Eq.~\ref{eq:reactance} with different values of $\delta$. Please note, we assumed here the ellipsoids to be prolate, hence, the x- and y-component of the polarizability are the same. This is in agreement with the assumptions further below. Elliptical nanoparticles with such shape can be obtained by various chemical methods \cite{Murphy_fabrication_elli,Foss_fabrication_elli}.

Equating the surface reactance for a given axis ratio to the surface reactance necessary to cancel the scattered field allows to identify the operational frequency for a given structure. It can be seen that the larger the aspect ratio of the ellipsoids the shorter the frequency where both reactances are the same. Changing the aspect ratio, therefore, is a suitable means to adjust the operational frequency upon request. To obtain insights into the qualitative scattering cross section of such a system, a volumetric homogenization method is used further below.

\subsection{Volumetric Homogenization}

Next, we consider the layer of plasmonic particles as a shell that is made from a homogeneous medium and which encloses the dielectric sphere to calculate the scattering cross section.

\begin{figure}[htbp]
\includegraphics[width=0.7\textwidth]{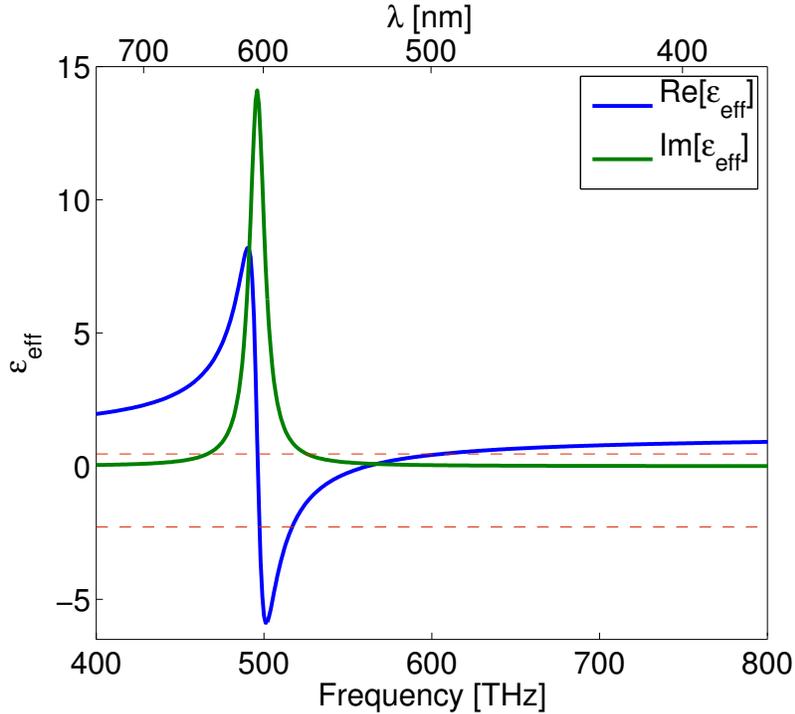}
  \caption{Effective relative permittivity of a material made of silver ellipsoids with semi axes $a_\textrm{x}=a_\textrm{y}=5$ nm and $a_\textrm{z}=23$ nm and a filling fraction of $f=0.065$. The dashed lines show the solutions of Eq.~\ref{cloaking_cond}.}
  \label{fig_eps_eff}
\end{figure}

\begin{figure}[htbp]
  \includegraphics[width=0.7\textwidth]{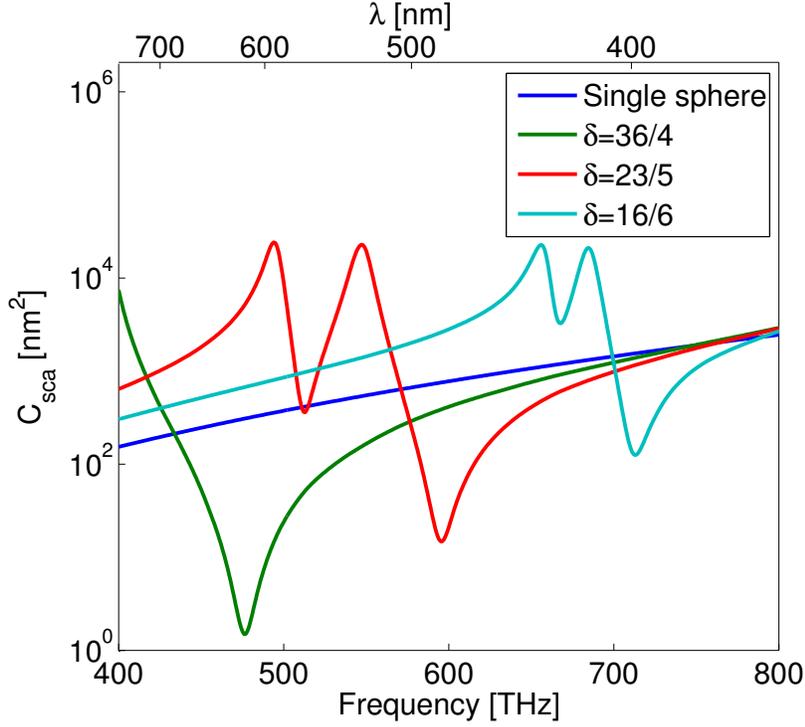}
  \caption{Scattering cross section of a single dielectric sphere with $\varepsilon_\textrm{sph}=2.1$ and the sphere covered with an effective material corresponding to different aspect ratios of the constituent ellipsoids.}
  \label{fig_ana}
\end{figure}

To assign an effective permittivity to the shell, we use the Clausius-Mossotti-relation for inclusions in a host medium with permittivity $\varepsilon_\textrm{h}$
\begin{align}
\varepsilon_\textrm{eff}(\omega)= \frac{\varepsilon_\textrm{h} +2/3 f \alpha(\omega)}{1-\frac{f\alpha(\omega)}{3\varepsilon_\textrm{h}}}~,
\label{CM}
\end{align}
where $f=0.065$ is the filling fraction of the particles in the host and $\alpha(\omega)$ is the polarizability of a single particle. The polarizability and host medium are the same as considered in the previous subsection, i.e.~we took again the spatial average of the three individual polarizabilities of the three axes.

To reduce the scattering of the system we need to satisfy the cloaking condition of a covered sphere \cite{alu_engheta_cloak}
\begin{align}
\gamma^3=\frac{(\varepsilon_\textrm{eff}(\omega)-\varepsilon_\textrm{b})(2\varepsilon_\textrm{eff}(\omega)+\varepsilon_\textrm{sph})}{(\varepsilon_\textrm{eff}(\omega)-\varepsilon_\textrm{sph})(2\varepsilon_\textrm{eff}(\omega)+\varepsilon_\textrm{b})}
\label{cloaking_cond} ~.
\end{align}
This equation is real valued for lossless core and background material, thus we only get a restriction on the real part of the cloak permittivity. We get two solutions for the chosen parameters: one positive $\Re[\varepsilon_\textrm{eff}(\omega_1)]=0.461$ and one negative $\Re[\varepsilon_\textrm{eff}(\omega_2)]=-2.278$, depicted as dashed lines in Fig.~\ref{fig_eps_eff}. The scattering of the coated sphere will be reduced at the positions where the real part of $\varepsilon_\textrm{eff}(\omega)$ is equal to those values. The remaining imaginary part of the permittivity slightly diminishes the cloaking effect, introducing absorption that is associated with additional forward scattering, due to the optical theorem \cite{fleury_absorption}.

Now, we can calculate the polarizability of the entire system by considering a sphere coated with a shell of the effective material. The polarizability of a coated sphere is given as \cite{Sihvola_shell}
\begin{align}
\alpha_\textrm{clo}(\omega)=3\frac{(\varepsilon_\textrm{eff}(\omega)-1)(\varepsilon_\textrm{sph}+2\varepsilon_\textrm{eff}(\omega))+\gamma^3(2\varepsilon_\textrm{eff}(\omega)+1)(\varepsilon_\textrm{sph}-\varepsilon_\textrm{eff}(\omega))}{(\varepsilon_\textrm{eff}(\omega)+2)(\varepsilon_\textrm{sph}+2\varepsilon_\textrm{eff}(\omega))+2\gamma^3(\varepsilon_\textrm{eff}(\omega)-1)(\varepsilon_\textrm{sph}-\varepsilon_\textrm{eff}(\omega))}~.
\end{align}
This can be used to calculate the scattering cross section of the system as \cite{bohren}
\begin{align}
C_\textrm{sca}(\omega)=k^4 r_\textrm{clo}^6 8\pi|{\alpha_\textrm{clo}(\omega)}|^2~.
\end{align}

As can be seen in Fig.~\ref{fig_ana}, the calculated scattering cross section of the covered sphere shows two peaks and two local minima. The first peak corresponds to the resonance position of the effective permittivity of the shell and the second corresponds to the plasmon resonance between the core and the shell, respectively. The minima correspond to the points where the cloaking condition of the covered sphere Eq.~\ref{cloaking_cond} is satisfied. A possible third minimum, is not observed. This corresponds to the spectral position where $\varepsilon_\textrm{eff}(\omega_3)=-2.278$ but which is directly in the resonance. There, the imaginary part of the effective permittivity is extremely high, such that the condition to suppress the scattering response cannot be met effectively. In fact, also the second (negative) solution does not lead to cloaking, because of the high losses near the resonance.

Comparable to the previous analytical method, we observe a high tunability of the operational frequency across the entire optical range that is considered. By changing the aspect ratio of the ellipsoids $\delta=a_\textrm{z}/a_\textrm{x}$, the effective permittivity of the shell material changes. This leads to different positions where the cloaking condition is satisfied. Otherwise the qualitative response is identical for all operational frequencies.

As we see, the predicted cloaking frequency, given by the intersection of the $X_\textrm{s}$ graphs with $X_\textrm{s}^\textrm{ideal}$, is in excellent agreement with the values from the volumetric homogenization (Fig.~\ref{fig_ana}).

It is interesting to note that, as it was also shown in Ref.~\onlinecite{monti_cloak}, volumetric homogenization provides quite accurate results even if the shell is rather thin and made of a single layer of nanoparticles. Further proofs of the applicability of the volumetric homogenization in such situation will be given in the results section.

We notice that we discussed there only the condition necessary to meet the positive solution. As we will see further below, this is the only solution that persists in the actual structure, where disorder possibly causes a degradation of the oscillator strength of the effective dispersion.

\section{Numerical Method}
\label{Numerical Method}

To be able to study the properties from an actual structure, we have to be able to simulate its optical response using full-wave numerical simulations. The structure itself constitutes in that sense a challenge, since it consists of a larger number of resonant objects with fine details. We are not able to take advantage of any periodicity or symmetry, since the elliptical nanoparticles shall be considered with a random orientation on top of the core object. This is the consequence from the bottom-up methods that we suggest to implement such scattering cancellation device. To cope with these challenges, we rely on an extension to the multiple scattering technique \cite{Xu_multiple_scattering,Muhlig_nanotips}. Our approach requires only information on the T-matrix of the isolated ellipsoid and the core sphere \cite{waterman_t-matrix,tsang1985theory,mishchenko}. This T-matrix can be obtained using multiple full-wave simulations of the optical response of the isolated object for different illumination schemes \cite{gimbutas_t-matrix}. The matrix contains generally the complete information on how the object interacts with electromagnetic radiation in the far and near field. However, for numerical calculations we need to truncate the T-matrix and take only the information up to a chosen multipole order into account.

We start by considering the optical response of a single object that is illuminated by a specific external electromagnetic field. The total field outside the object can always be decomposed into the incident and the scattered field. With respect to a central coordinate of the scattering object, these fields can be expanded into vector spherical harmonics $\vN_{nm}^{(l)}(r,\theta,\phi)$ and $\vM_{nm}^{(l)}(r,\theta,\phi)$, the functions with indices $l=1,2$ contain Bessel functions of first and second kind, respectively, and indices $l=3,4$ correspond to Hankel functions of first and second kind, respectively. The exact choice of the function requires further physical reasoning. For the scattered electric field that has to satisfy the Sommerfeld radiation condition the expansion reads as
\begin{align}
\vE_{\mathrm{sca}}(\textbf{r},\omega) =\sum_{n=1}^{\infty}\sum_{m=-n}^{n}k^{2}E_{nm} \left[a_{nm}(\omega)\vN_{nm}^{(3)}(\textbf{r},\omega)+b_{nm}(\omega)\vM_{nm}^{(3)}(\textbf{r},\omega)\right].
\label{eq:scattered}
\end{align}
For the incident field that has to be finite in the center of the coordinate systems the expansion reads as
\begin{align}
\vE_{\mathrm{inc}}(\textbf{r},\omega) =-\sum_{n=1}^{\infty}\sum_{m=-n}^{n}k^{2}E_{nm} \left[p_{nm}(\omega)\vN_{nm}^{(1)}(\textbf{r},\omega)+q_{nm}(\omega)\vM_{nm}^{(1)}(\textbf{r},\omega)\right].
\end{align}
$k^{2}=\E(\omega)\M(\omega)\left(\omega^{2}/c^{2}\right)$ is the dispersion relation in the surrounding medium characterized by the permittivity $\E(\omega)$ and the permeability $\M(\omega)$ and 
\begin{align}
E_{nm}=|\vE_\mathrm{inc}|i^n(2n+1)\frac{(n-m)!}{(n+m)!}
\end{align}
is a scaling factor with $|\vE_\mathrm{inc}|$ being the magnitude of the incident field.
The expansion coefficients $a_{nm}(\omega)$ and $b_{nm}(\omega)$ are called scattering coefficients. The expansion coefficients $p_{nm}(\omega)$ and $q_{nm}(\omega)$ describe the illumination. They are analytically known for a plane wave or a Gaussian beam, but they can be also kept entirely free to expand an arbitrary illumination close to the scattering object. 

The link between the scattering coefficients and the incident coefficients is given by the T-matrix of the object. This matrix contains the entire information on how a specific objects responds to an external electromagnetic field and reads as
\begin{align}
\left(\begin{matrix} \textbf{a} \\ \textbf{b} \end{matrix}\right)=\textbf{T} \. \left(\begin{matrix} \textbf{p} \\ \textbf{q} \end{matrix}\right),
\label{eq:t-matrix}
\end{align}
$\textbf{a}$ and $\textbf{b}$ are concatenated vectors containing the scattering coefficients of the outgoing wave and $\textbf{p}$ and $\textbf{q}$ are concatenated vectors containing the coefficients of the incident wave. The T-matrix links them. 

The T-matrix is analytically known only for canonical objects, e.g. a sphere. In the case of a sphere, the T-matrix is diagonal and the values on the diagonal are known as the Mie coefficients. This reflects the fact that vector spherical harmonics with different quantum numbers are orthogonal. This orthogonality is not broken by a spherical object. However, for an arbitrary scatterer the T-matrix is dense in general.
Using Eq.~\ref{eq:t-matrix}, we can calculate the T-matrix for any given scatterer. Specifically, we use an existing full wave code to illuminate the scatterer with plane waves from different directions and calculate the scattered fields. 
Then, we can calculate sets of incident coefficients $\textbf{p}$, $\textbf{q}$ and corresponding scattering coefficients $\textbf{a}$, $\textbf{b}$. These are calculated by projecting the scattered and illuminating fields onto vector spherical harmonics
\begin{align}
a_{nm}(\omega)&=\frac{\int\limits_{0}^{2\pi}\! \int\limits_{0}^{\pi}\! \vE_{\mathrm{sca}}(r=R,\theta,\phi,\omega)\vN_{nm}^*(r=R,\theta,\phi,\omega)\sin{\theta} \, \mathrm{d}\theta \mathrm{d}\phi}{k^2 E_{nm}\int\limits_{0}^{2\pi}\!\int\limits_{0}^{\pi}\!|\vN_{nm}(r=R,\theta,\phi,\omega)|^2 \sin{\theta} \,\mathrm{d}\theta \mathrm{d}\phi}~,\nonumber \\
b_{nm}(\omega)&=\frac{\int\limits_{0}^{2\pi}\! \int\limits_{0}^{\pi}\! \vE_{\mathrm{sca}}(r=R,\theta,\phi,\omega)\vM_{nm}^*(r=R,\theta,\phi,\omega)\sin{\theta} \, \mathrm{d}\theta \mathrm{d}\phi}{k^2 E_{nm}\int\limits_{0}^{2\pi}\!\int\limits_{0}^{\pi}\!|\vM_{nm}(r=R,\theta,\phi,\omega)|^2 \sin{\theta} \,\mathrm{d}\theta \mathrm{d}\phi}~,
\end{align}
where $R$ is a fixed radius of a sphere containing the object at which the fields are evaluated \cite{bohren}. Having the incident and scattering coefficients available, the calculation of $\textbf{T}$ is done by inverting the system of equations. In order to have a solvable system, we need to consider at least $2N(N+2)$ different illuminations, which is the dimension of the T-matrix. Here $N$ is the multipolar order we want to take into account, where $N=1$ corresponds to dipoles, $N=2$ to quadrupoles and so on.
Using Eq.~\ref{eq:t-matrix} we can calculate the scattering coefficients of any particle for any given incident field, represented by the incident field coefficients. Specifically, we apply this procedure to quantify the exact scattering properties of the ellipsoidal nanoparticles.

Now, instead of just a single isolated scatterer, we consider a cluster of particles. In the multiple scattering algorithm, the illumination to each particle is written as a superposition of the external illumination and the scattered field from all other spheres. The incident field on the $j^{\mathrm{th}}$ sphere expanded in its local coordinate system is given by
\begin{align}
\vE_{\mathrm{inc}}^j(\textbf{r},\omega)=\vE_{\mathrm{inc}}(\textbf{r},\omega)+\sum_{i\neq j}\vE_{\mathrm{sca}}(\textbf{r},\omega)(i,j)~,
\label{eq:self}
\end{align}
where $(i, j)$ denotes the transformation from the $i^{\mathrm{th}}$ to the $j^{\mathrm{th}}$ coordinate system. This is done by translating the vector spherical harmonics with the translational addition theorems
\begin{align}
\vN_{nm}^{(3)}(\textbf{r},\omega) &=\sum_{n'=1}^{\infty}\sum_{m'=-n'}^{n'}\left[A0_{nm}^{n'm'}(i,j)\vN_{n'm'}^{(1)}(\textbf{r}',\omega)+B0_{nm}^{n'm'}(i,j)\vM_{n'm'}^{(1)}(\textbf{r}',\omega)\right], \nonumber \\
\vM_{nm}^{(3)}(\textbf{r},\omega) &=\sum_{n'=1}^{\infty}\sum_{m'=-n'}^{n'}\left[A0_{nm}^{n'm'}(i,j)\vM_{n'm'}^{(1)}(\textbf{r}',\omega)+B0_{nm}^{n'm'}(i,j)\vN_{n'm'}^{(1)}(\textbf{r}',\omega)\right].
\label{eq:transl}
\end{align}
The translation coefficients $A0_{n'm'}^{nm}(i,j)$ and $B0_{n'm'}^{nm}(i,j)$ can be found in Ref.~\onlinecite{Stein_1961}.
Then the T-matrix transforms as
\begin{align}
\textbf{T}(j)=\left(\begin{matrix} \textbf{A} & \textbf{B} \\ \textbf{B} & \textbf{A}\end{matrix}\right) ^{\! \! *} \! \! (i,j) \cdot \textbf{T}(i) \cdot \left(\begin{matrix} \textbf{A} & \textbf{B} \\ \textbf{B} & \textbf{A}\end{matrix}\right) \! \! (i,j) 
\end{align}
Here $\textbf{A}$ and $\textbf{B}$ form a matrix of translation coefficients that can be obtained from
\begin{align}
A_{nm}^{n'm'}=\frac{E_{n'm'}}{E_{nm}} A0_{nm}^{n'm'}.
\end{align}

For non-spherical particles however, not only the position but also the orientation is important to determine the scattering. This can be accounted for by applying a rotation to the T-matrix \cite{mishchenko}
\begin{align}
T_{nmn'm'}^{\mathrm{rot}}=\sum_{m_1=-n}^{n} \sum_{m_2=-n'}^{n'}\left[D_{m'm_2}^{n'}(\alpha,\beta,\gamma)\right]^* T_{nm_1n'm_2} D_{mm_1}^n(\alpha,\beta,\gamma)~.
\label{eq:rot}
\end{align}
Here the Wigner D-matrix
\begin{align}
D_{ml}^{n}(\alpha,\beta,\gamma)=e^{-im\alpha}d_{ml}^n(\beta)e^{-il\gamma}
\end{align}
depends on the Euler angles $\alpha,\beta,\gamma$ of the rotation and the Wigner d-function is given as
\begin{align}
d_{ml}^{n}(\beta)=A_{ml}^n (1-\cos\beta)^{(m-l)/2}(1+\cos\beta)^{-(m+l)/2}\mathrm{\partial}_{\cos\beta}^{n-l}\left[(1-\cos\beta)^{n-m}(1+\cos\beta)^{n+m}\right]
\end{align}
with the factor
\begin{align}
A_{ml}^n=\frac{(-1)^{n-l}}{2^n}\sqrt{\frac{(n+l)!}{(n-m)!(n+m)!(n-l)!}}~.
\end{align}

By combining Eqs.~\ref{eq:t-matrix}, \ref{eq:self}, \ref{eq:transl}, and \ref{eq:rot} we arrive at a system of equations that can be solved self-consistently for the scattering coefficients of all constitutive particles with index $j$ under the illumination of the external field and the scattered field of all other particles
\begin{align}
\left(\begin{matrix} \textbf{a} \\ \textbf{b} \end{matrix}\right)^{\! \! (j)}=\textbf{T}^{\mathrm{rot}} \. \left( \left(\begin{matrix} \textbf{p} \\ \textbf{q} \end{matrix}\right)^{\! \! (j)}- \sum_{j\neq i} \left(\left(\begin{matrix} \textbf{a} \\ \textbf{b} \end{matrix}\right)^{\! \! (i)} \. \left(\begin{matrix} \textbf{A} \\ \textbf{B} \end{matrix}\right) \! \! (i,j)  \right) \right).
\label{eq:multimie}
\end{align}

The scattering coefficients of the whole cluster can finally be obtained by translating all coefficients back to the central coordinate system and summing them together
\begin{align}
\left(\begin{matrix} \textbf{a} \\ \textbf{b} \end{matrix}\right)=\sum_{j=1}^J \left(\begin{matrix} \textbf{a} \\ \textbf{b} \end{matrix}\right)^{\! \! (j)} \. \left(\begin{matrix} \textbf{A} \\ \textbf{B} \end{matrix}\right) \! \! (j,1) ~.
\label{eq:coeff}
\end{align}
From these scattering coefficients all further quantities can be calculated.

This formalism speeds up considerably the calculation of the optical response from a large ensemble of scattering objects. To calculate the scattering coefficients of an array of specially or randomly oriented objects, the T-matrix has to be calculated only once and can be used for each constituent by translating and rotating it by the desired amount. This is done using addition theorems for the vector spherical harmonics \cite{Stein_1961,mishchenko}. With this method we are able to simulate the electromagnetic response of ellipsoids arbitrarily arranged at the surface of a larger dielectric sphere to investigate the cloaking in more detail. In order to perform numerical calculations, we have to truncate the infinite sums, e.~g.~in Eq.~\ref{eq:scattered} at a finite number $N$ which represents the multipolar order. This means contributions from orders up to $N$ are taken into account.

\section{Results}
\label{Results}
\begin{figure}[htbp]

  \raisebox{-0.2\height}{\includegraphics[width=0.13\textwidth]{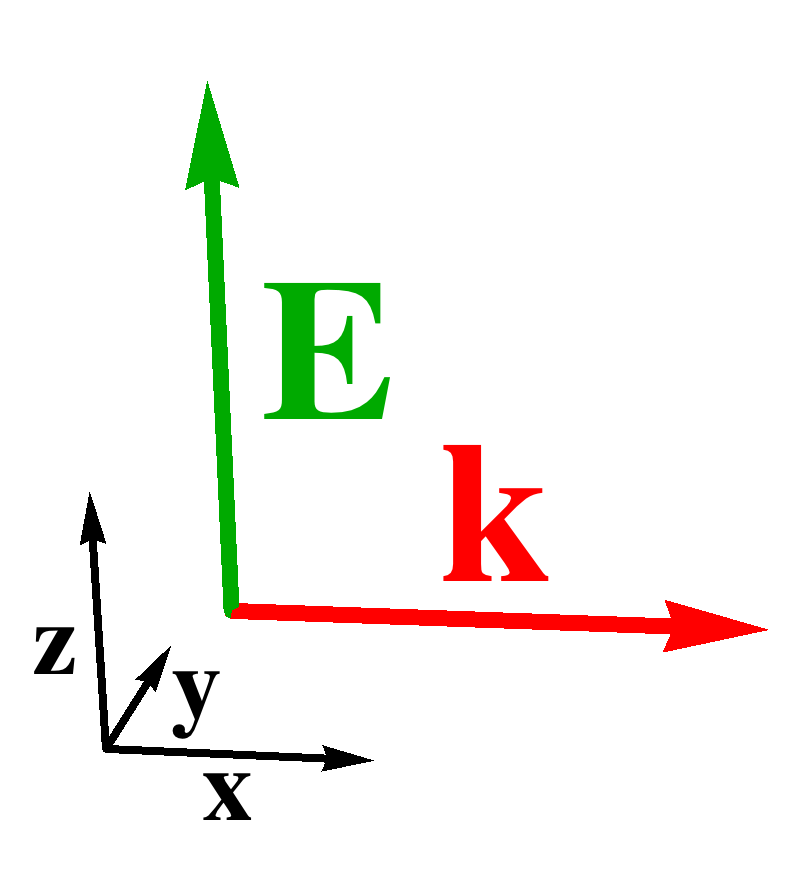}}
  \subfloat[Completely ordered \ \ \ \ and aligned.]{\raisebox{-0.5\height}{\includegraphics[width=0.28\textwidth]{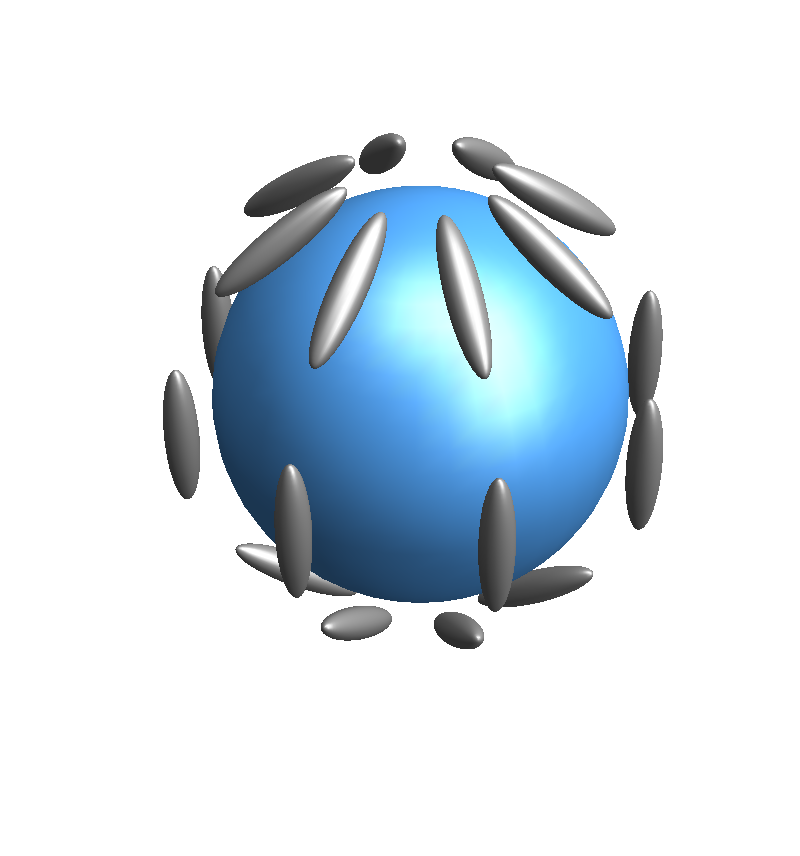}}\label{figuabpP01a}} 
  \subfloat[Random distribution, \ \ \ but same alignment.]{\raisebox{-0.5\height}{\includegraphics[width=0.28\textwidth]{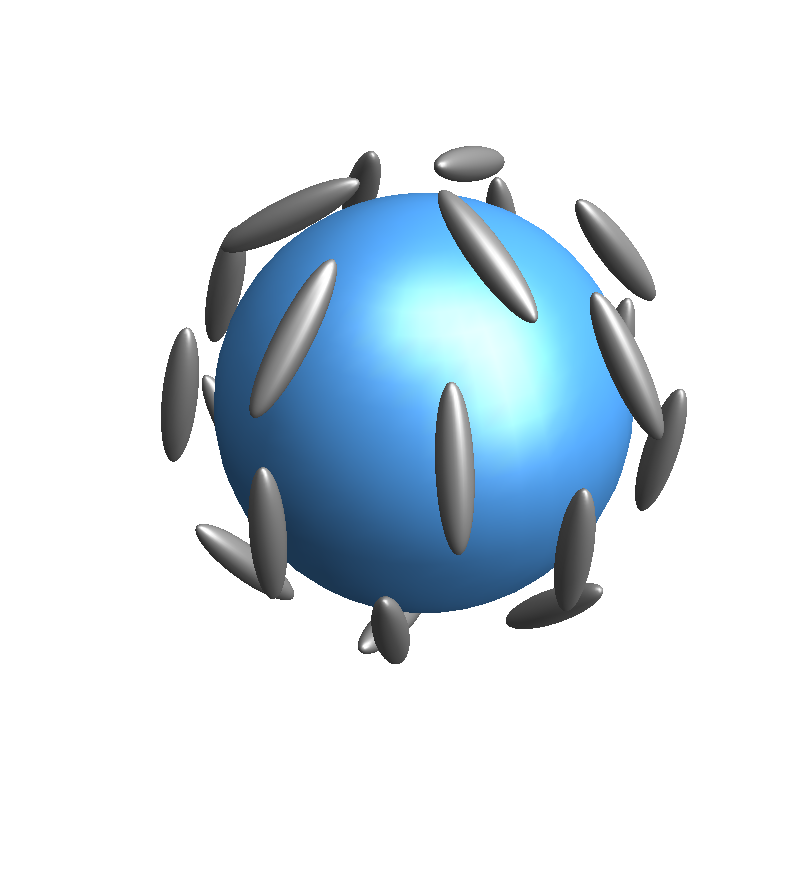}}\label{figuabpP01b}}
  \subfloat[Random distribution \ \ and random alignment.]{\raisebox{-0.5\height}{\includegraphics[width=0.28\textwidth]{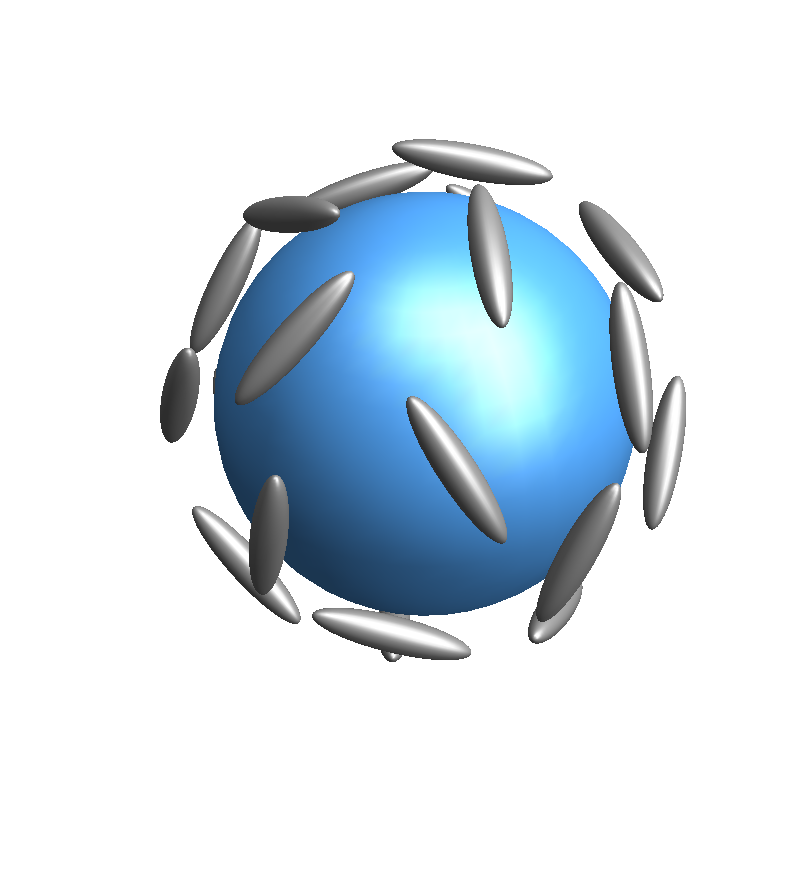}}\label{figuabpP01c}}
  \caption{Different distributions and orientations of silver ellipsoids on a dielectric sphere. The illumination direction and polarization are depicted on the left hand side.}
  \label{fig_distributions}
\end{figure}
\begin{figure}[htbp]
\centering \includegraphics[width=0.75\textwidth]{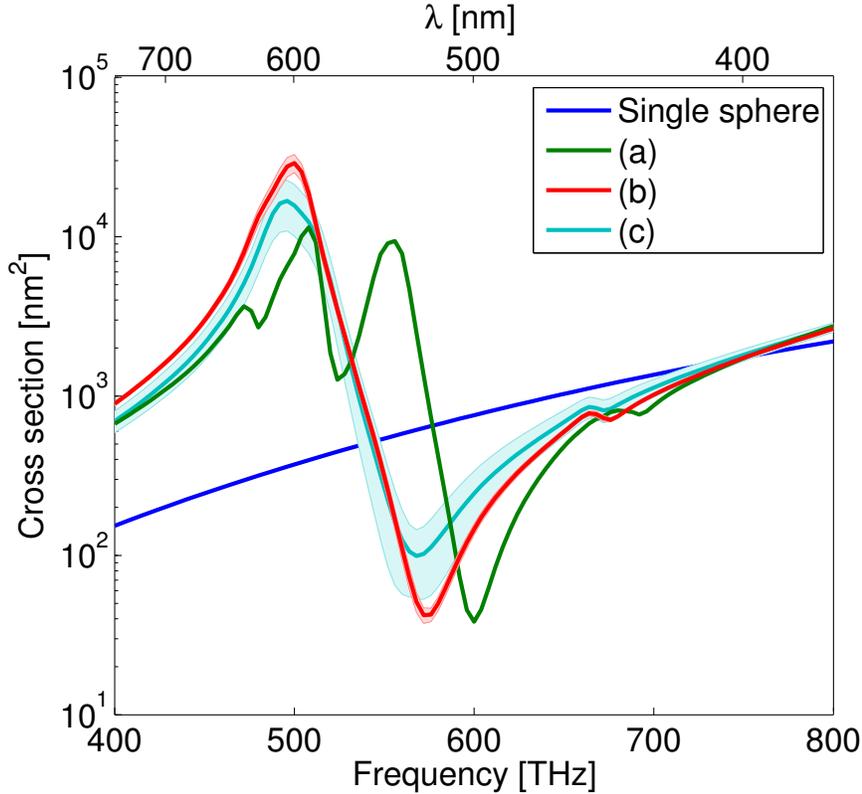} \caption{Numerically obtained scattering cross sections of different systems corresponding to the geometries shown in Fig.~\ref{fig_distributions}: (a) Completely ordered and aligned, (b) random distribution, same alignment, (c) random distribution, random alignment. For (b) and (c) we display the averaged cross section of $100$ simulations with different realizations of the random geometry, the shaded region shows the standard deviation of the sample.
The ellipsoids have the semi axes of $a_\textrm{x}=a_\textrm{y}=5$ nm and $a_\textrm{z}=23$ nm.}
\label{fig_cloaking_n}
\end{figure}
We consider now a dielectric sphere with $\varepsilon_\textrm{sph}=2.1$ and a radius of $r_\textrm{Sph}=61$ nm covered with $24$ silver ellipsoids. For the permittivity of the nanoparticles $\varepsilon_\textrm{p}(\omega)$ we use established experimental data \cite{johnson_christy}.
With the proposed T-matrix algorithm it is possible to calculate the scattering response from different distributions and orientations of the silver ellipsoids on the central sphere in short time and with high precision. We truncate the infinite sums at $N=3$. This means we take dipoles, quadrupoles and octopoles into account, however, already the quadupole contribution is almost negligible due to the small size of the structure compared to the wavelength.

In Fig.~\ref{fig_distributions}, three configurations are shown that shall be further investigated. Figure~\ref{fig_distributions}(a) shows a completely deterministic ordering of the ellipsoids. They have defined distances and are lying tangent to the surface of the core sphere with an orientation that maximizes the projection of the long axis onto the polarization direction of the incident field. This was done to enhance the polarizability of the shell. However, when this nanostructure is fabricated with self assembly techniques, we have no direct control over the exact position of the ellipsoids. To reflect this we show in Fig.~\ref{fig_distributions}(b) random distributions of nanoparticles on the surface of the core sphere only adhering to a minimal distance to ensure that the ellipsoids are not in contact with each other. The particles are, however, still aligned with respect to the incident field polarization like geometry (a). This could be realized for example by applying an external field during the self assembly process to force the particles to align \cite{park_align}. Finally, Fig.~\ref{fig_distributions}(c) shows the same random distribution additionally with random orientation of the ellipsoids on the surface of the sphere.

As shown in Fig.~\ref{fig_cloaking_n}, the three different geometries result in similar qualitative cloaking behavior. The scattering cross section is reduced by almost one order of magnitude around the desired frequency of $600$ THz as anticipated by the analytic calculations. 
However, there are a few notable observations. Firstly, as expected, the scattering cancellation is weaker for the case of randomly oriented particles (c), because less ellipsoids are aligned along the polarization of the electric field with their large semi axis $a_\textrm{z}$. Furthermore the second dip of the scattering cross section at approximately $530$ THz, which was predicted by analytical considerations, vanishes if we introduce disorder. This can be attributed to the fact that the effective permittivity of the shell shows a weaker dispersion, i.e. the oscillator strength of the effective Lorentzian-like dispersion is reduced. The effective permittivity of the disordered shell material no longer attains those values corresponding to the negative solution of the cloaking condition Eq.~\ref{cloaking_cond}. Therefore, only the positive solution persists. Finally, the frequency of minimal scattering is shifted towards the red part of the spectrum when we introduce disorder. This can likewise be explained with a lower dispersion of $\varepsilon_\textrm{eff}(\omega)$ of the shell, because the value of the positive solution of Eq.~\ref{cloaking_cond} is attained at a lower frequency. This can be understood intuitively by imagining a weaker dispersive $\varepsilon_\textrm{eff}(\omega)$ in  Fig.~\ref{fig_eps_eff}.

However, because of the overall similarity of the results at the target frequency, we can state that the design is rather robust to uncertainties, which may arise in fabrication. We wish to stress that similar implementations of nominal identical geometries have nearly identical scattering responses, as is evidenced by the small red-shaded area in Fig.~\ref{fig_cloaking_n}, which shows the standard deviation for different random positions of the ellipsoids. The exact details of the implementation of a specific disordered structure are not important. Variation of the orientation of the ellipsoids has a slightly larger impact on the scattering reduction, as demonstrated by the cyan-shaded area. But the scattering cross section is still significantly reduced, note that this means that the cloaking behavior of the disordered geometry (c) is generally independent of the direction and polarization of the incident plane wave illumination.

\begin{figure}[htbp]
\centering \includegraphics[width=0.75\textwidth]{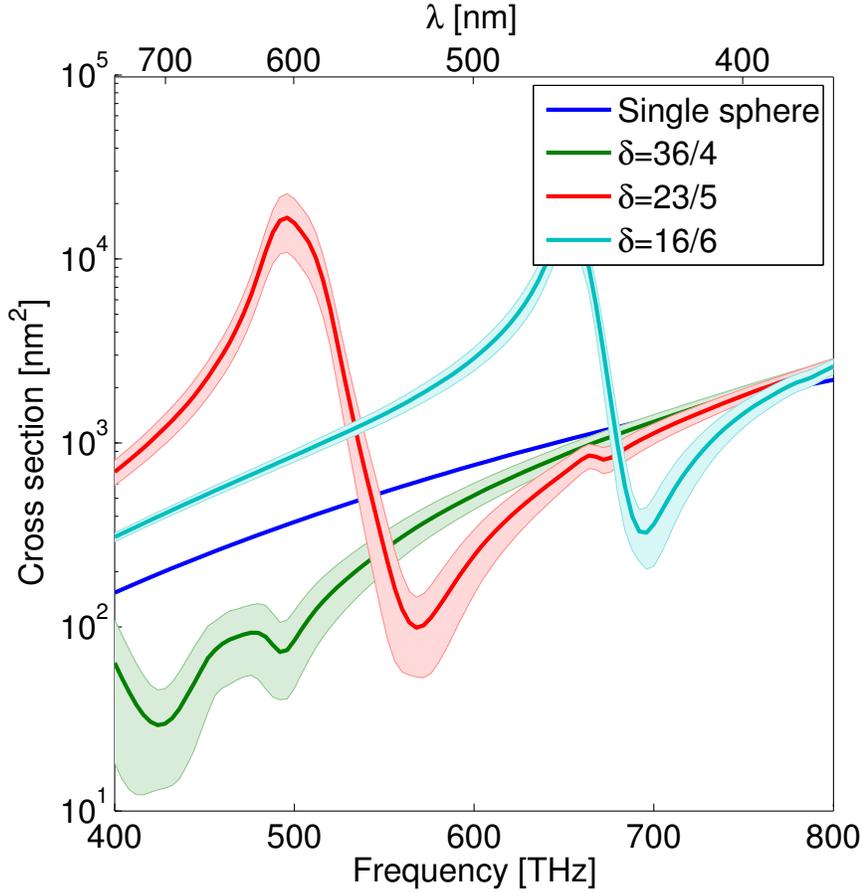} \caption{Numerically obtained scattering cross sections of spheres decorated with ellipsoids with different aspect ratios. All systems have the same disordered geometry shown in Fig.~\ref{fig_distributions}(c). The thick lines show the average of $100$ different random realizations and the shaded region depicts the standard deviation}
\label{fig_cloaking_aspect}
\end{figure}

We showed in section \ref{Theoretical considerations} that the desired cloaking frequency can be tuned across a large fraction of the visible electromagnetic spectrum. This is achieved by changing the aspect ratio $\delta$ of the metallic ellipsoids, while keeping the volume constant. Now, we perform full wave calculations with the method established in section \ref{Numerical Method}. As can be seen in Fig.~\ref{fig_cloaking_aspect} the cloaking frequency ranges from $430$ THz to $700$ THz if we change the aspect ratio. An important point to consider is that the aspect ratios do not reach extreme values and are entirely in the range of what is possible to fabricate experimentally \cite{Murphy_fabrication_elli}.
The cloaking is slightly distorted in the case of the high aspect ratio ellipsoids. This is due to the fact that the particles are very long and are almost touching. This can cause strong coupling which is completely excluded from the theoretical considerations. However, the scattering cross section is still significantly reduced.

In a final step we wish to explicitly compare the different methods that can be used to describe the functionality of the device. As shown in section \ref{Theoretical considerations}, the cloak can be described as a homogeneous shell with an effective permittivity $\eps_\textrm{eff}(\omega)$. Now, we not only want to do this analytically by considering the analytical expression for the polarizability of an ellipsoid but we wish to do such analysis numerically. For this purpose, we calculated the polarizability $\alpha(\omega)$ of a single ellipsoid numerically, by using the T-matrix of a single ellipsoid.
The polarizability can be obtained from the scattering coefficients of the first order \cite{muhlig_multipole}. For example, for an illumination with z-polarized light the polarizability is given as
\begin{align}
\alpha_\textrm{z} (\omega)= -\sqrt{12\pi}i Z_0 k \cdot a_{10}(\omega)~.
\end{align}

\begin{figure}[htbp]
\centering \includegraphics[width=0.7\textwidth]{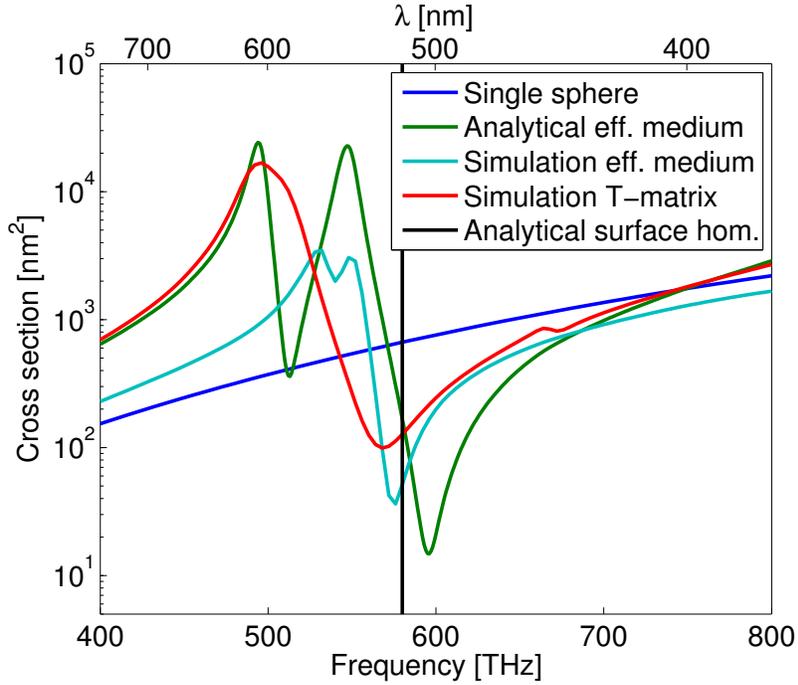} \caption{Scattering cross section of the single sphere, analytically obtained effective medium coating, the numerical calculation of the effective shell, the complete full wave simulation of geometry (c) and the predicted cloaking frequency of the surface homogenization.}
\label{fig_cloaking_all}
\end{figure}

Then, we take Eq.~\ref{CM} to get the effective permittivity of the homogenized shell and conduct a full wave simulation of a sphere covered with the effective material.
As shown in Fig.~\ref{fig_cloaking_all} the functional behavior of the analytical calculation with volume homogenization and the numerical simulation are very similar. However, the scattering reduction is weaker. This can be attributed to the lower dispersion in the effective permittivity. The dispersion is weaker, because additionally radiative losses are automatically included in the full wave solutions. In contrast, in the analytical discussion of the polarizability of the ellipsoid (Eq.~\ref{analytical_eiipsoid}) this has not been considered. The equation remains only strictly valid in the quasi-static regime. Additionally, the negative solution of Eq.~\ref{cloaking_cond} appears at higher frequencies. This is due to a shift in the resonance of $\varepsilon_\textrm{eff}$ and also the weaker dispersion.
In summary and beyond all these detailed insights, we can state that we get very good agreement in the cloaking position predicted by the different numerical and analytic methods. However, it is important to note that, especially in the visible, intrinsic absorption of the constituents from which the shell is made can be large if plasmonic particles are used. This might lead to an enhancement of the total extinction cross-section of the particle, even though the scattering cross section is reduced at the operational frequency \cite{Miller_comment}.

\section{Conclusion}
We introduced a new design for a scattering cancellation cloak. Instead of using plasmonic spheres or complicated metasurfaces we decorate a dielectric sphere with silver ellipsoids. The main advantage is the high tunability that makes it possible to shift the cloaking frequency across the optical domain by changing the aspect ratio of the ellipsoids. Additionally, the design is feasible for fabrication with self-assembly techniques because the particles can be fabricated chemically and the design is robust to changes of the distribution.
\\ 

Furthermore, we outlined a powerful and versatile numerical method that can be used to simulate the scattering of a large ensemble of arbitrary particles. With the procedure at hand we just calculate the T-matrix of an object once with the help of available full wave solvers and use it to calculate the scattering properties of larger clusters.  We calculated the scattering of different realizations of the cloaking geometry. This was used to validate two different approaches to describe the scattering cancellation cloak analytically. We showed that all methods had a very good agreement, especially in predicting the cloaking frequency.
The ellipsoids that are used in the shell are very well tunable and allow for cloaking at specific frequencies over the entire visible spectrum, by changing the aspect ratio in an isotropic geometry.

\section*{Acknowledgement}
We wish to thank Vassilios Yannopapas for his help with the T-matrix rotation algorithm.
This work was supported by the German Science Foundation within project RO 3640/4-1. M.F.~also acknowledges the support by the Karlsruhe School of Optics \& Photonics (KSOP).


\begin{thebibliography}{46}%
\makeatletter
\providecommand \@ifxundefined [1]{%
 \@ifx{#1\undefined}
}%
\providecommand \@ifnum [1]{%
 \ifnum #1\expandafter \@firstoftwo
 \else \expandafter \@secondoftwo
 \fi
}%
\providecommand \@ifx [1]{%
 \ifx #1\expandafter \@firstoftwo
 \else \expandafter \@secondoftwo
 \fi
}%
\providecommand \natexlab [1]{#1}%
\providecommand \enquote  [1]{``#1''}%
\providecommand \bibnamefont  [1]{#1}%
\providecommand \bibfnamefont [1]{#1}%
\providecommand \citenamefont [1]{#1}%
\providecommand \href@noop [0]{\@secondoftwo}%
\providecommand \href [0]{\begingroup \@sanitize@url \@href}%
\providecommand \@href[1]{\@@startlink{#1}\@@href}%
\providecommand \@@href[1]{\endgroup#1\@@endlink}%
\providecommand \@sanitize@url [0]{\catcode `\\12\catcode `\$12\catcode
  `\&12\catcode `\#12\catcode `\^12\catcode `\_12\catcode `\%12\relax}%
\providecommand \@@startlink[1]{}%
\providecommand \@@endlink[0]{}%
\providecommand \url  [0]{\begingroup\@sanitize@url \@url }%
\providecommand \@url [1]{\endgroup\@href {#1}{\urlprefix }}%
\providecommand \urlprefix  [0]{URL }%
\providecommand \Eprint [0]{\href }%
\providecommand \doibase [0]{http://dx.doi.org/}%
\providecommand \selectlanguage [0]{\@gobble}%
\providecommand \bibinfo  [0]{\@secondoftwo}%
\providecommand \bibfield  [0]{\@secondoftwo}%
\providecommand \translation [1]{[#1]}%
\providecommand \BibitemOpen [0]{}%
\providecommand \bibitemStop [0]{}%
\providecommand \bibitemNoStop [0]{.\EOS\space}%
\providecommand \EOS [0]{\spacefactor3000\relax}%
\providecommand \BibitemShut  [1]{\csname bibitem#1\endcsname}%
\let\auto@bib@innerbib\@empty
\bibitem [{\citenamefont {Evangelou}\ \emph {et~al.}(2012)\citenamefont
  {Evangelou}, \citenamefont {Yannopapas},\ and\ \citenamefont
  {Paspalakis}}]{Yannopapas_transparency}%
  \BibitemOpen
  \bibfield  {author} {\bibinfo {author} {\bibfnamefont {S.}~\bibnamefont
  {Evangelou}}, \bibinfo {author} {\bibfnamefont {V.}~\bibnamefont
  {Yannopapas}}, \ and\ \bibinfo {author} {\bibfnamefont {E.}~\bibnamefont
  {Paspalakis}},\ }\href@noop {} {\bibfield  {journal} {\bibinfo  {journal}
  {Physical Review A}\ }\textbf {\bibinfo {volume} {86}},\ \bibinfo {pages}
  {053811} (\bibinfo {year} {2012})}\BibitemShut {NoStop}%
\bibitem [{\citenamefont {Knight}\ \emph {et~al.}(2011)\citenamefont {Knight},
  \citenamefont {Sobhani}, \citenamefont {Nordlander},\ and\ \citenamefont
  {Halas}}]{Nordlander_nanoantennas}%
  \BibitemOpen
  \bibfield  {author} {\bibinfo {author} {\bibfnamefont {M.~W.}\ \bibnamefont
  {Knight}}, \bibinfo {author} {\bibfnamefont {H.}~\bibnamefont {Sobhani}},
  \bibinfo {author} {\bibfnamefont {P.}~\bibnamefont {Nordlander}}, \ and\
  \bibinfo {author} {\bibfnamefont {N.~J.}\ \bibnamefont {Halas}},\ }\href@noop
  {} {\bibfield  {journal} {\bibinfo  {journal} {Science}\ }\textbf {\bibinfo
  {volume} {332}},\ \bibinfo {pages} {702} (\bibinfo {year}
  {2011})}\BibitemShut {NoStop}%
\bibitem [{\citenamefont {Gansel}\ \emph {et~al.}(2009)\citenamefont {Gansel},
  \citenamefont {Thiel}, \citenamefont {Rill}, \citenamefont {Decker},
  \citenamefont {Bade}, \citenamefont {Saile}, \citenamefont {von Freymann},
  \citenamefont {Linden},\ and\ \citenamefont {Wegener}}]{Wegener_helix}%
  \BibitemOpen
  \bibfield  {author} {\bibinfo {author} {\bibfnamefont {J.~K.}\ \bibnamefont
  {Gansel}}, \bibinfo {author} {\bibfnamefont {M.}~\bibnamefont {Thiel}},
  \bibinfo {author} {\bibfnamefont {M.~S.}\ \bibnamefont {Rill}}, \bibinfo
  {author} {\bibfnamefont {M.}~\bibnamefont {Decker}}, \bibinfo {author}
  {\bibfnamefont {K.}~\bibnamefont {Bade}}, \bibinfo {author} {\bibfnamefont
  {V.}~\bibnamefont {Saile}}, \bibinfo {author} {\bibfnamefont
  {G.}~\bibnamefont {von Freymann}}, \bibinfo {author} {\bibfnamefont
  {S.}~\bibnamefont {Linden}}, \ and\ \bibinfo {author} {\bibfnamefont
  {M.}~\bibnamefont {Wegener}},\ }\href@noop {} {\bibfield  {journal} {\bibinfo
   {journal} {Science}\ }\textbf {\bibinfo {volume} {325}},\ \bibinfo {pages}
  {1513} (\bibinfo {year} {2009})}\BibitemShut {NoStop}%
\bibitem [{\citenamefont {Sadecka}\ \emph
  {et~al.}(2015{\natexlab{a}})\citenamefont {Sadecka}, \citenamefont {Gajc},
  \citenamefont {Orlinski}, \citenamefont {Surma}, \citenamefont {Klos},
  \citenamefont {Jozwik-Biala}, \citenamefont {Sobczak}, \citenamefont
  {Dluzewski}, \citenamefont {Toudert},\ and\ \citenamefont {Pawlak}}]{Pawlak}%
  \BibitemOpen
  \bibfield  {author} {\bibinfo {author} {\bibfnamefont {K.}~\bibnamefont
  {Sadecka}}, \bibinfo {author} {\bibfnamefont {M.}~\bibnamefont {Gajc}},
  \bibinfo {author} {\bibfnamefont {K.}~\bibnamefont {Orlinski}}, \bibinfo
  {author} {\bibfnamefont {H.~B.}\ \bibnamefont {Surma}}, \bibinfo {author}
  {\bibfnamefont {A.}~\bibnamefont {Klos}}, \bibinfo {author} {\bibfnamefont
  {I.}~\bibnamefont {Jozwik-Biala}}, \bibinfo {author} {\bibfnamefont
  {K.}~\bibnamefont {Sobczak}}, \bibinfo {author} {\bibfnamefont
  {P.}~\bibnamefont {Dluzewski}}, \bibinfo {author} {\bibfnamefont
  {J.}~\bibnamefont {Toudert}}, \ and\ \bibinfo {author} {\bibfnamefont
  {D.~A.}\ \bibnamefont {Pawlak}},\ }\href@noop {} {\bibfield  {journal}
  {\bibinfo  {journal} {Advanced Optical Materials}\ }\textbf {\bibinfo
  {volume} {3}},\ \bibinfo {pages} {381} (\bibinfo {year}
  {2015}{\natexlab{a}})}\BibitemShut {NoStop}%
\bibitem [{\citenamefont {Sadecka}\ \emph
  {et~al.}(2015{\natexlab{b}})\citenamefont {Sadecka}, \citenamefont {Toudert},
  \citenamefont {Surma},\ and\ \citenamefont {Pawlak}}]{Pawlak2}%
  \BibitemOpen
  \bibfield  {author} {\bibinfo {author} {\bibfnamefont {K.}~\bibnamefont
  {Sadecka}}, \bibinfo {author} {\bibfnamefont {J.}~\bibnamefont {Toudert}},
  \bibinfo {author} {\bibfnamefont {H.~B.}\ \bibnamefont {Surma}}, \ and\
  \bibinfo {author} {\bibfnamefont {D.~A.}\ \bibnamefont {Pawlak}},\
  }\href@noop {} {\bibfield  {journal} {\bibinfo  {journal} {Optics Express}\
  }\textbf {\bibinfo {volume} {23}},\ \bibinfo {pages} {19098} (\bibinfo {year}
  {2015}{\natexlab{b}})}\BibitemShut {NoStop}%
\bibitem [{\citenamefont {Pendry}\ \emph {et~al.}(2006)\citenamefont {Pendry},
  \citenamefont {Schurig},\ and\ \citenamefont {Smith}}]{pendry_cloak}%
  \BibitemOpen
  \bibfield  {author} {\bibinfo {author} {\bibfnamefont {J.~B.}\ \bibnamefont
  {Pendry}}, \bibinfo {author} {\bibfnamefont {D.}~\bibnamefont {Schurig}}, \
  and\ \bibinfo {author} {\bibfnamefont {D.~R.}\ \bibnamefont {Smith}},\
  }\href@noop {} {\bibfield  {journal} {\bibinfo  {journal} {Science}\ }\textbf
  {\bibinfo {volume} {312}},\ \bibinfo {pages} {1780} (\bibinfo {year}
  {2006})}\BibitemShut {NoStop}%
\bibitem [{\citenamefont {Leonhardt}(2006)}]{Leonhardt_cloak}%
  \BibitemOpen
  \bibfield  {author} {\bibinfo {author} {\bibfnamefont {U.}~\bibnamefont
  {Leonhardt}},\ }\href@noop {} {\bibfield  {journal} {\bibinfo  {journal}
  {Science}\ }\textbf {\bibinfo {volume} {312}},\ \bibinfo {pages} {1777}
  (\bibinfo {year} {2006})}\BibitemShut {NoStop}%
\bibitem [{\citenamefont {Jiang}\ \emph {et~al.}(2008)\citenamefont {Jiang},
  \citenamefont {Cui}, \citenamefont {Yang}, \citenamefont {Cheng},
  \citenamefont {Liu},\ and\ \citenamefont {Smith}}]{Jiang_cloak}%
  \BibitemOpen
  \bibfield  {author} {\bibinfo {author} {\bibfnamefont {W.~X.}\ \bibnamefont
  {Jiang}}, \bibinfo {author} {\bibfnamefont {T.~J.}\ \bibnamefont {Cui}},
  \bibinfo {author} {\bibfnamefont {X.~M.}\ \bibnamefont {Yang}}, \bibinfo
  {author} {\bibfnamefont {Q.}~\bibnamefont {Cheng}}, \bibinfo {author}
  {\bibfnamefont {R.}~\bibnamefont {Liu}}, \ and\ \bibinfo {author}
  {\bibfnamefont {D.~R.}\ \bibnamefont {Smith}},\ }\href@noop {} {\bibfield
  {journal} {\bibinfo  {journal} {Applied Physics Letters}\ }\textbf {\bibinfo
  {volume} {93}},\ \bibinfo {pages} {194102} (\bibinfo {year}
  {2008})}\BibitemShut {NoStop}%
\bibitem [{\citenamefont {Schumann}\ \emph {et~al.}(2015)\citenamefont
  {Schumann}, \citenamefont {Wiesendanger}, \citenamefont {Goldschmidt},
  \citenamefont {Bl{\"a}si}, \citenamefont {Bittkau}, \citenamefont {Paetzold},
  \citenamefont {Sprafke}, \citenamefont {Wehrspohn}, \citenamefont
  {Rockstuhl},\ and\ \citenamefont {Wegener}}]{Schumann_cloaking_solar_cell}%
  \BibitemOpen
  \bibfield  {author} {\bibinfo {author} {\bibfnamefont {M.~F.}\ \bibnamefont
  {Schumann}}, \bibinfo {author} {\bibfnamefont {S.}~\bibnamefont
  {Wiesendanger}}, \bibinfo {author} {\bibfnamefont {J.~C.}\ \bibnamefont
  {Goldschmidt}}, \bibinfo {author} {\bibfnamefont {B.}~\bibnamefont
  {Bl{\"a}si}}, \bibinfo {author} {\bibfnamefont {K.}~\bibnamefont {Bittkau}},
  \bibinfo {author} {\bibfnamefont {U.~W.}\ \bibnamefont {Paetzold}}, \bibinfo
  {author} {\bibfnamefont {A.}~\bibnamefont {Sprafke}}, \bibinfo {author}
  {\bibfnamefont {R.~B.}\ \bibnamefont {Wehrspohn}}, \bibinfo {author}
  {\bibfnamefont {C.}~\bibnamefont {Rockstuhl}}, \ and\ \bibinfo {author}
  {\bibfnamefont {M.}~\bibnamefont {Wegener}},\ }\href@noop {} {\bibfield
  {journal} {\bibinfo  {journal} {Optica}\ }\textbf {\bibinfo {volume} {2}},\
  \bibinfo {pages} {850} (\bibinfo {year} {2015})}\BibitemShut {NoStop}%
\bibitem [{\citenamefont {Li}\ and\ \citenamefont
  {Pendry}(2008)}]{Pendry_carpet}%
  \BibitemOpen
  \bibfield  {author} {\bibinfo {author} {\bibfnamefont {J.}~\bibnamefont
  {Li}}\ and\ \bibinfo {author} {\bibfnamefont {J.~B.}\ \bibnamefont
  {Pendry}},\ }\href@noop {} {\bibfield  {journal} {\bibinfo  {journal}
  {Physical Review Letters}\ }\textbf {\bibinfo {volume} {101}},\ \bibinfo
  {pages} {203901} (\bibinfo {year} {2008})}\BibitemShut {NoStop}%
\bibitem [{\citenamefont {Valentine}\ \emph {et~al.}(2009)\citenamefont
  {Valentine}, \citenamefont {Li}, \citenamefont {Zentgraf}, \citenamefont
  {Bartal},\ and\ \citenamefont {Zhang}}]{Zentgraf_carpet}%
  \BibitemOpen
  \bibfield  {author} {\bibinfo {author} {\bibfnamefont {J.}~\bibnamefont
  {Valentine}}, \bibinfo {author} {\bibfnamefont {J.}~\bibnamefont {Li}},
  \bibinfo {author} {\bibfnamefont {T.}~\bibnamefont {Zentgraf}}, \bibinfo
  {author} {\bibfnamefont {G.}~\bibnamefont {Bartal}}, \ and\ \bibinfo {author}
  {\bibfnamefont {X.}~\bibnamefont {Zhang}},\ }\href@noop {} {\bibfield
  {journal} {\bibinfo  {journal} {Nature Materials}\ }\textbf {\bibinfo
  {volume} {8}},\ \bibinfo {pages} {568} (\bibinfo {year} {2009})}\BibitemShut
  {NoStop}%
\bibitem [{\citenamefont {Ergin}\ \emph {et~al.}(2011)\citenamefont {Ergin},
  \citenamefont {Fischer},\ and\ \citenamefont {Wegener}}]{Wegener_carpet}%
  \BibitemOpen
  \bibfield  {author} {\bibinfo {author} {\bibfnamefont {T.}~\bibnamefont
  {Ergin}}, \bibinfo {author} {\bibfnamefont {J.}~\bibnamefont {Fischer}}, \
  and\ \bibinfo {author} {\bibfnamefont {M.}~\bibnamefont {Wegener}},\
  }\href@noop {} {\bibfield  {journal} {\bibinfo  {journal} {Physical Review
  Letters}\ }\textbf {\bibinfo {volume} {107}},\ \bibinfo {pages} {173901}
  (\bibinfo {year} {2011})}\BibitemShut {NoStop}%
\bibitem [{\citenamefont {Lee}\ \emph {et~al.}(2009)\citenamefont {Lee},
  \citenamefont {Blair}, \citenamefont {Tamma}, \citenamefont {Wu},
  \citenamefont {Rhee}, \citenamefont {Summers},\ and\ \citenamefont
  {Park}}]{park_carpet}%
  \BibitemOpen
  \bibfield  {author} {\bibinfo {author} {\bibfnamefont {J.~H.}\ \bibnamefont
  {Lee}}, \bibinfo {author} {\bibfnamefont {J.}~\bibnamefont {Blair}}, \bibinfo
  {author} {\bibfnamefont {V.~A.}\ \bibnamefont {Tamma}}, \bibinfo {author}
  {\bibfnamefont {Q.}~\bibnamefont {Wu}}, \bibinfo {author} {\bibfnamefont
  {S.~J.}\ \bibnamefont {Rhee}}, \bibinfo {author} {\bibfnamefont {C.~J.}\
  \bibnamefont {Summers}}, \ and\ \bibinfo {author} {\bibfnamefont
  {W.}~\bibnamefont {Park}},\ }\href@noop {} {\bibfield  {journal} {\bibinfo
  {journal} {Optics Express}\ }\textbf {\bibinfo {volume} {17}},\ \bibinfo
  {pages} {12922} (\bibinfo {year} {2009})}\BibitemShut {NoStop}%
\bibitem [{\citenamefont {Al{\`u}}\ and\ \citenamefont
  {Engheta}(2005)}]{alu_engheta_cloak}%
  \BibitemOpen
  \bibfield  {author} {\bibinfo {author} {\bibfnamefont {A.}~\bibnamefont
  {Al{\`u}}}\ and\ \bibinfo {author} {\bibfnamefont {N.}~\bibnamefont
  {Engheta}},\ }\href@noop {} {\bibfield  {journal} {\bibinfo  {journal}
  {Physical Review E}\ }\textbf {\bibinfo {volume} {72}},\ \bibinfo {pages}
  {016623} (\bibinfo {year} {2005})}\BibitemShut {NoStop}%
\bibitem [{\citenamefont {Silveirinha}\ \emph {et~al.}(2008)\citenamefont
  {Silveirinha}, \citenamefont {Al{\`u}},\ and\ \citenamefont
  {Engheta}}]{Silveirinha_cloak}%
  \BibitemOpen
  \bibfield  {author} {\bibinfo {author} {\bibfnamefont {M.~G.}\ \bibnamefont
  {Silveirinha}}, \bibinfo {author} {\bibfnamefont {A.}~\bibnamefont
  {Al{\`u}}}, \ and\ \bibinfo {author} {\bibfnamefont {N.}~\bibnamefont
  {Engheta}},\ }\href@noop {} {\bibfield  {journal} {\bibinfo  {journal}
  {Physical Review B}\ }\textbf {\bibinfo {volume} {78}},\ \bibinfo {pages}
  {075107} (\bibinfo {year} {2008})}\BibitemShut {NoStop}%
\bibitem [{\citenamefont {Bilotti}\ \emph {et~al.}(2008)\citenamefont
  {Bilotti}, \citenamefont {Tricarico},\ and\ \citenamefont
  {Vegni}}]{Bilotti_te_tm}%
  \BibitemOpen
  \bibfield  {author} {\bibinfo {author} {\bibfnamefont {F.}~\bibnamefont
  {Bilotti}}, \bibinfo {author} {\bibfnamefont {S.}~\bibnamefont {Tricarico}},
  \ and\ \bibinfo {author} {\bibfnamefont {L.}~\bibnamefont {Vegni}},\
  }\href@noop {} {\bibfield  {journal} {\bibinfo  {journal} {New Journal of
  Physics}\ }\textbf {\bibinfo {volume} {10}},\ \bibinfo {pages} {115035}
  (\bibinfo {year} {2008})}\BibitemShut {NoStop}%
\bibitem [{\citenamefont {Monti}\ \emph
  {et~al.}(2015{\natexlab{a}})\citenamefont {Monti}, \citenamefont {Al{\`u}},
  \citenamefont {Toscano},\ and\ \citenamefont {Bilotti}}]{monti_cloak1}%
  \BibitemOpen
  \bibfield  {author} {\bibinfo {author} {\bibfnamefont {A.}~\bibnamefont
  {Monti}}, \bibinfo {author} {\bibfnamefont {A.}~\bibnamefont {Al{\`u}}},
  \bibinfo {author} {\bibfnamefont {A.}~\bibnamefont {Toscano}}, \ and\
  \bibinfo {author} {\bibfnamefont {F.}~\bibnamefont {Bilotti}},\ }\href@noop
  {} {\bibfield  {journal} {\bibinfo  {journal} {Journal of Applied Physics}\
  }\textbf {\bibinfo {volume} {117}},\ \bibinfo {pages} {123103} (\bibinfo
  {year} {2015}{\natexlab{a}})}\BibitemShut {NoStop}%
\bibitem [{\citenamefont {Monti}\ \emph
  {et~al.}(2015{\natexlab{b}})\citenamefont {Monti}, \citenamefont {Al{\`u}},
  \citenamefont {Toscano},\ and\ \citenamefont {Bilotti}}]{monti_cloak2}%
  \BibitemOpen
  \bibfield  {author} {\bibinfo {author} {\bibfnamefont {A.}~\bibnamefont
  {Monti}}, \bibinfo {author} {\bibfnamefont {A.}~\bibnamefont {Al{\`u}}},
  \bibinfo {author} {\bibfnamefont {A.}~\bibnamefont {Toscano}}, \ and\
  \bibinfo {author} {\bibfnamefont {F.}~\bibnamefont {Bilotti}},\ }in\
  \href@noop {} {\emph {\bibinfo {booktitle} {Photonics}}},\ Vol.~\bibinfo
  {volume} {2}\ (\bibinfo {organization} {Multidisciplinary Digital Publishing
  Institute},\ \bibinfo {year} {2015})\ pp.\ \bibinfo {pages}
  {540--552}\BibitemShut {NoStop}%
\bibitem [{\citenamefont {Kerker}(1975)}]{Kerker_invisibility}%
  \BibitemOpen
  \bibfield  {author} {\bibinfo {author} {\bibfnamefont {M.}~\bibnamefont
  {Kerker}},\ }\href@noop {} {\bibfield  {journal} {\bibinfo  {journal}
  {Journal of the Optical Society of America}\ }\textbf {\bibinfo {volume}
  {65}},\ \bibinfo {pages} {376} (\bibinfo {year} {1975})}\BibitemShut
  {NoStop}%
\bibitem [{\citenamefont {Rockstuhl}\ and\ \citenamefont
  {Scharf}(2013)}]{Rockstuhl_book}%
  \BibitemOpen
  \bibfield  {author} {\bibinfo {author} {\bibfnamefont {C.}~\bibnamefont
  {Rockstuhl}}\ and\ \bibinfo {author} {\bibfnamefont {T.}~\bibnamefont
  {Scharf}},\ }\href@noop {} {\emph {\bibinfo {title} {Amorphous
  Nanophotonics}}}\ (\bibinfo  {publisher} {Springer Science \& Business
  Media},\ \bibinfo {year} {2013})\BibitemShut {NoStop}%
\bibitem [{\citenamefont {Onal}\ and\ \citenamefont
  {Guven}(2015)}]{Onal_antenna_cloak}%
  \BibitemOpen
  \bibfield  {author} {\bibinfo {author} {\bibfnamefont {E.~D.}\ \bibnamefont
  {Onal}}\ and\ \bibinfo {author} {\bibfnamefont {K.}~\bibnamefont {Guven}},\
  }\href@noop {} {\bibfield  {journal} {\bibinfo  {journal} {arXiv preprint
  arXiv:1511.01312}\ } (\bibinfo {year} {2015})}\BibitemShut {NoStop}%
\bibitem [{\citenamefont {Bilotti}\ \emph {et~al.}(2011)\citenamefont
  {Bilotti}, \citenamefont {Tricarico}, \citenamefont {Pierini},\ and\
  \citenamefont {Vegni}}]{Bilotti_tip_cloak}%
  \BibitemOpen
  \bibfield  {author} {\bibinfo {author} {\bibfnamefont {F.}~\bibnamefont
  {Bilotti}}, \bibinfo {author} {\bibfnamefont {S.}~\bibnamefont {Tricarico}},
  \bibinfo {author} {\bibfnamefont {F.}~\bibnamefont {Pierini}}, \ and\
  \bibinfo {author} {\bibfnamefont {L.}~\bibnamefont {Vegni}},\ }\href@noop {}
  {\bibfield  {journal} {\bibinfo  {journal} {Optics Letters}\ }\textbf
  {\bibinfo {volume} {36}},\ \bibinfo {pages} {211} (\bibinfo {year}
  {2011})}\BibitemShut {NoStop}%
\bibitem [{\citenamefont {Tricarico}\ \emph {et~al.}(2010)\citenamefont
  {Tricarico}, \citenamefont {Bilotti},\ and\ \citenamefont
  {Vegni}}]{Bilotti_force_cloak}%
  \BibitemOpen
  \bibfield  {author} {\bibinfo {author} {\bibfnamefont {S.}~\bibnamefont
  {Tricarico}}, \bibinfo {author} {\bibfnamefont {F.}~\bibnamefont {Bilotti}},
  \ and\ \bibinfo {author} {\bibfnamefont {L.}~\bibnamefont {Vegni}},\
  }\href@noop {} {\bibfield  {journal} {\bibinfo  {journal} {Physical Review
  B}\ }\textbf {\bibinfo {volume} {82}},\ \bibinfo {pages} {045109} (\bibinfo
  {year} {2010})}\BibitemShut {NoStop}%
\bibitem [{\citenamefont {Al{\`u}}(2009)}]{alu_invisibility_prb}%
  \BibitemOpen
  \bibfield  {author} {\bibinfo {author} {\bibfnamefont {A.}~\bibnamefont
  {Al{\`u}}},\ }\href@noop {} {\bibfield  {journal} {\bibinfo  {journal}
  {Physical Peview B}\ }\textbf {\bibinfo {volume} {80}},\ \bibinfo {pages}
  {245115} (\bibinfo {year} {2009})}\BibitemShut {NoStop}%
\bibitem [{\citenamefont {Rainwater}\ \emph {et~al.}(2012)\citenamefont
  {Rainwater}, \citenamefont {Kerkhoff}, \citenamefont {Melin}, \citenamefont
  {Soric}, \citenamefont {Moreno},\ and\ \citenamefont
  {Al{\`u}}}]{Alu_experimental_mantle_cloak}%
  \BibitemOpen
  \bibfield  {author} {\bibinfo {author} {\bibfnamefont {D.}~\bibnamefont
  {Rainwater}}, \bibinfo {author} {\bibfnamefont {A.}~\bibnamefont {Kerkhoff}},
  \bibinfo {author} {\bibfnamefont {K.}~\bibnamefont {Melin}}, \bibinfo
  {author} {\bibfnamefont {J.}~\bibnamefont {Soric}}, \bibinfo {author}
  {\bibfnamefont {G.}~\bibnamefont {Moreno}}, \ and\ \bibinfo {author}
  {\bibfnamefont {A.}~\bibnamefont {Al{\`u}}},\ }\href@noop {} {\bibfield
  {journal} {\bibinfo  {journal} {New Journal of Physics}\ }\textbf {\bibinfo
  {volume} {14}},\ \bibinfo {pages} {013054} (\bibinfo {year}
  {2012})}\BibitemShut {NoStop}%
\bibitem [{\citenamefont {M{\"u}hlig}\ \emph
  {et~al.}(2011{\natexlab{a}})\citenamefont {M{\"u}hlig}, \citenamefont
  {Farhat}, \citenamefont {Rockstuhl},\ and\ \citenamefont
  {Lederer}}]{Muhlig_prb_cloak}%
  \BibitemOpen
  \bibfield  {author} {\bibinfo {author} {\bibfnamefont {S.}~\bibnamefont
  {M{\"u}hlig}}, \bibinfo {author} {\bibfnamefont {M.}~\bibnamefont {Farhat}},
  \bibinfo {author} {\bibfnamefont {C.}~\bibnamefont {Rockstuhl}}, \ and\
  \bibinfo {author} {\bibfnamefont {F.}~\bibnamefont {Lederer}},\ }\href@noop
  {} {\bibfield  {journal} {\bibinfo  {journal} {Physical Review B}\ }\textbf
  {\bibinfo {volume} {83}},\ \bibinfo {pages} {195116} (\bibinfo {year}
  {2011}{\natexlab{a}})}\BibitemShut {NoStop}%
\bibitem [{\citenamefont {Muehlig}\ \emph {et~al.}(2013)\citenamefont
  {Muehlig}, \citenamefont {Cunningham}, \citenamefont {Dintinger},
  \citenamefont {Farhat}, \citenamefont {Bin~Hasan}, \citenamefont {Scharf},
  \citenamefont {Buergi}, \citenamefont {Lederer},\ and\ \citenamefont
  {Rockstuhl}}]{Muhlig_selfassembled_cloak}%
  \BibitemOpen
  \bibfield  {author} {\bibinfo {author} {\bibfnamefont {S.}~\bibnamefont
  {Muehlig}}, \bibinfo {author} {\bibfnamefont {A.}~\bibnamefont {Cunningham}},
  \bibinfo {author} {\bibfnamefont {J.}~\bibnamefont {Dintinger}}, \bibinfo
  {author} {\bibfnamefont {M.}~\bibnamefont {Farhat}}, \bibinfo {author}
  {\bibfnamefont {S.}~\bibnamefont {Bin~Hasan}}, \bibinfo {author}
  {\bibfnamefont {T.}~\bibnamefont {Scharf}}, \bibinfo {author} {\bibfnamefont
  {T.}~\bibnamefont {Buergi}}, \bibinfo {author} {\bibfnamefont
  {F.}~\bibnamefont {Lederer}}, \ and\ \bibinfo {author} {\bibfnamefont
  {C.}~\bibnamefont {Rockstuhl}},\ }\href@noop {} {\bibfield  {journal}
  {\bibinfo  {journal} {Scientific Reports}\ }\textbf {\bibinfo {volume} {3}},\
  \bibinfo {pages} {189353} (\bibinfo {year} {2013})}\BibitemShut {NoStop}%
\bibitem [{\citenamefont {Waterman}(1965)}]{waterman_t-matrix}%
  \BibitemOpen
  \bibfield  {author} {\bibinfo {author} {\bibfnamefont {P.~C.}\ \bibnamefont
  {Waterman}},\ }\href@noop {} {\bibfield  {journal} {\bibinfo  {journal}
  {Proceedings of the IEEE}\ }\textbf {\bibinfo {volume} {53}},\ \bibinfo
  {pages} {805} (\bibinfo {year} {1965})}\BibitemShut {NoStop}%
\bibitem [{\citenamefont {Saeidi}\ and\ \citenamefont {van~der
  Weide}(2013)}]{saeidi_surface}%
  \BibitemOpen
  \bibfield  {author} {\bibinfo {author} {\bibfnamefont {C.}~\bibnamefont
  {Saeidi}}\ and\ \bibinfo {author} {\bibfnamefont {D.}~\bibnamefont {van~der
  Weide}},\ }\href@noop {} {\bibfield  {journal} {\bibinfo  {journal} {Optics
  Express}\ }\textbf {\bibinfo {volume} {21}},\ \bibinfo {pages} {16170}
  (\bibinfo {year} {2013})}\BibitemShut {NoStop}%
\bibitem [{\citenamefont {Tretyakov}(2003)}]{tretyakov_book}%
  \BibitemOpen
  \bibfield  {author} {\bibinfo {author} {\bibfnamefont {S.}~\bibnamefont
  {Tretyakov}},\ }\href@noop {} {\emph {\bibinfo {title} {Analytical modeling
  in applied electromagnetics}}}\ (\bibinfo  {publisher} {Artech House},\
  \bibinfo {year} {2003})\BibitemShut {NoStop}%
\bibitem [{\citenamefont {Bohren}\ and\ \citenamefont
  {Huffman}(2008)}]{bohren}%
  \BibitemOpen
  \bibfield  {author} {\bibinfo {author} {\bibfnamefont {C.~F.}\ \bibnamefont
  {Bohren}}\ and\ \bibinfo {author} {\bibfnamefont {D.~R.}\ \bibnamefont
  {Huffman}},\ }\href@noop {} {\emph {\bibinfo {title} {Absorption and
  scattering of light by small particles}}}\ (\bibinfo  {publisher} {John Wiley
  \& Sons},\ \bibinfo {year} {2008})\BibitemShut {NoStop}%
\bibitem [{\citenamefont {Johnson}\ and\ \citenamefont
  {Christy}(1972)}]{johnson_christy}%
  \BibitemOpen
  \bibfield  {author} {\bibinfo {author} {\bibfnamefont {P.~B.}\ \bibnamefont
  {Johnson}}\ and\ \bibinfo {author} {\bibfnamefont {R.-W.}\ \bibnamefont
  {Christy}},\ }\href@noop {} {\bibfield  {journal} {\bibinfo  {journal}
  {Physical Peview B}\ }\textbf {\bibinfo {volume} {6}},\ \bibinfo {pages}
  {4370} (\bibinfo {year} {1972})}\BibitemShut {NoStop}%
\bibitem [{\citenamefont {Murphy}\ \emph {et~al.}(2005)\citenamefont {Murphy},
  \citenamefont {Sau}, \citenamefont {Gole}, \citenamefont {Orendorff},
  \citenamefont {Gao}, \citenamefont {Gou}, \citenamefont {Hunyadi},\ and\
  \citenamefont {Li}}]{Murphy_fabrication_elli}%
  \BibitemOpen
  \bibfield  {author} {\bibinfo {author} {\bibfnamefont {C.~J.}\ \bibnamefont
  {Murphy}}, \bibinfo {author} {\bibfnamefont {T.~K.}\ \bibnamefont {Sau}},
  \bibinfo {author} {\bibfnamefont {A.~M.}\ \bibnamefont {Gole}}, \bibinfo
  {author} {\bibfnamefont {C.~J.}\ \bibnamefont {Orendorff}}, \bibinfo {author}
  {\bibfnamefont {J.}~\bibnamefont {Gao}}, \bibinfo {author} {\bibfnamefont
  {L.}~\bibnamefont {Gou}}, \bibinfo {author} {\bibfnamefont {S.~E.}\
  \bibnamefont {Hunyadi}}, \ and\ \bibinfo {author} {\bibfnamefont
  {T.}~\bibnamefont {Li}},\ }\href@noop {} {\bibfield  {journal} {\bibinfo
  {journal} {The Journal of Physical Chemistry B}\ }\textbf {\bibinfo {volume}
  {109}},\ \bibinfo {pages} {13857} (\bibinfo {year} {2005})}\BibitemShut
  {NoStop}%
\bibitem [{\citenamefont {Foss~Jr}\ \emph {et~al.}(1994)\citenamefont
  {Foss~Jr}, \citenamefont {Hornyak}, \citenamefont {Stockert},\ and\
  \citenamefont {Martin}}]{Foss_fabrication_elli}%
  \BibitemOpen
  \bibfield  {author} {\bibinfo {author} {\bibfnamefont {C.~A.}\ \bibnamefont
  {Foss~Jr}}, \bibinfo {author} {\bibfnamefont {G.~L.}\ \bibnamefont
  {Hornyak}}, \bibinfo {author} {\bibfnamefont {J.~A.}\ \bibnamefont
  {Stockert}}, \ and\ \bibinfo {author} {\bibfnamefont {C.~R.}\ \bibnamefont
  {Martin}},\ }\href@noop {} {\bibfield  {journal} {\bibinfo  {journal} {The
  Journal of Physical Chemistry}\ }\textbf {\bibinfo {volume} {98}},\ \bibinfo
  {pages} {2963} (\bibinfo {year} {1994})}\BibitemShut {NoStop}%
\bibitem [{\citenamefont {Fleury}\ \emph {et~al.}(2014)\citenamefont {Fleury},
  \citenamefont {Soric},\ and\ \citenamefont {Al{\`u}}}]{fleury_absorption}%
  \BibitemOpen
  \bibfield  {author} {\bibinfo {author} {\bibfnamefont {R.}~\bibnamefont
  {Fleury}}, \bibinfo {author} {\bibfnamefont {J.}~\bibnamefont {Soric}}, \
  and\ \bibinfo {author} {\bibfnamefont {A.}~\bibnamefont {Al{\`u}}},\
  }\href@noop {} {\bibfield  {journal} {\bibinfo  {journal} {Physical Review
  B}\ }\textbf {\bibinfo {volume} {89}},\ \bibinfo {pages} {045122} (\bibinfo
  {year} {2014})}\BibitemShut {NoStop}%
\bibitem [{\citenamefont {Sihvola}\ and\ \citenamefont
  {Lindell}(1989)}]{Sihvola_shell}%
  \BibitemOpen
  \bibfield  {author} {\bibinfo {author} {\bibfnamefont {A.}~\bibnamefont
  {Sihvola}}\ and\ \bibinfo {author} {\bibfnamefont {I.}~\bibnamefont
  {Lindell}},\ }\href@noop {} {\bibfield  {journal} {\bibinfo  {journal}
  {Journal of Electromagnetic Waves and Applications}\ }\textbf {\bibinfo
  {volume} {3}},\ \bibinfo {pages} {37} (\bibinfo {year} {1989})}\BibitemShut
  {NoStop}%
\bibitem [{\citenamefont {Monti}\ \emph {et~al.}(2011)\citenamefont {Monti},
  \citenamefont {Bilotti},\ and\ \citenamefont {Toscano}}]{monti_cloak}%
  \BibitemOpen
  \bibfield  {author} {\bibinfo {author} {\bibfnamefont {A.}~\bibnamefont
  {Monti}}, \bibinfo {author} {\bibfnamefont {F.}~\bibnamefont {Bilotti}}, \
  and\ \bibinfo {author} {\bibfnamefont {A.}~\bibnamefont {Toscano}},\
  }\href@noop {} {\bibfield  {journal} {\bibinfo  {journal} {Optics Letters}\
  }\textbf {\bibinfo {volume} {36}},\ \bibinfo {pages} {4479} (\bibinfo {year}
  {2011})}\BibitemShut {NoStop}%
\bibitem [{\citenamefont {Xu}(1995)}]{Xu_multiple_scattering}%
  \BibitemOpen
  \bibfield  {author} {\bibinfo {author} {\bibfnamefont {Y.-l.}\ \bibnamefont
  {Xu}},\ }\href@noop {} {\bibfield  {journal} {\bibinfo  {journal} {Applied
  Optics}\ }\textbf {\bibinfo {volume} {34}},\ \bibinfo {pages} {4573}
  (\bibinfo {year} {1995})}\BibitemShut {NoStop}%
\bibitem [{\citenamefont {M{\"u}hlig}\ \emph {et~al.}(2010)\citenamefont
  {M{\"u}hlig}, \citenamefont {Rockstuhl}, \citenamefont {Pniewski},
  \citenamefont {Simovski}, \citenamefont {Tretyakov},\ and\ \citenamefont
  {Lederer}}]{Muhlig_nanotips}%
  \BibitemOpen
  \bibfield  {author} {\bibinfo {author} {\bibfnamefont {S.}~\bibnamefont
  {M{\"u}hlig}}, \bibinfo {author} {\bibfnamefont {C.}~\bibnamefont
  {Rockstuhl}}, \bibinfo {author} {\bibfnamefont {J.}~\bibnamefont {Pniewski}},
  \bibinfo {author} {\bibfnamefont {C.~R.}\ \bibnamefont {Simovski}}, \bibinfo
  {author} {\bibfnamefont {S.~A.}\ \bibnamefont {Tretyakov}}, \ and\ \bibinfo
  {author} {\bibfnamefont {F.}~\bibnamefont {Lederer}},\ }\href@noop {}
  {\bibfield  {journal} {\bibinfo  {journal} {Physical Review B}\ }\textbf
  {\bibinfo {volume} {81}},\ \bibinfo {pages} {075317} (\bibinfo {year}
  {2010})}\BibitemShut {NoStop}%
\bibitem [{\citenamefont {Tsang}\ \emph {et~al.}(1985)\citenamefont {Tsang},
  \citenamefont {Kong},\ and\ \citenamefont {Shin}}]{tsang1985theory}%
  \BibitemOpen
  \bibfield  {author} {\bibinfo {author} {\bibfnamefont {L.}~\bibnamefont
  {Tsang}}, \bibinfo {author} {\bibfnamefont {J.}~\bibnamefont {Kong}}, \ and\
  \bibinfo {author} {\bibfnamefont {R.}~\bibnamefont {Shin}},\ }\href@noop {}
  {\emph {\bibinfo {title} {Theory of microwave remote sensing}}}\ (\bibinfo
  {publisher} {Wiley-Interscience},\ \bibinfo {year} {1985})\BibitemShut
  {NoStop}%
\bibitem [{\citenamefont {Mishchenko}\ \emph {et~al.}(1996)\citenamefont
  {Mishchenko}, \citenamefont {Travis},\ and\ \citenamefont
  {Mackowski}}]{mishchenko}%
  \BibitemOpen
  \bibfield  {author} {\bibinfo {author} {\bibfnamefont {M.~I.}\ \bibnamefont
  {Mishchenko}}, \bibinfo {author} {\bibfnamefont {L.~D.}\ \bibnamefont
  {Travis}}, \ and\ \bibinfo {author} {\bibfnamefont {D.~W.}\ \bibnamefont
  {Mackowski}},\ }\href@noop {} {\bibfield  {journal} {\bibinfo  {journal}
  {Journal of Quantitative Spectroscopy and Radiative Transfer}\ }\textbf
  {\bibinfo {volume} {55}},\ \bibinfo {pages} {535} (\bibinfo {year}
  {1996})}\BibitemShut {NoStop}%
\bibitem [{\citenamefont {Gimbutas}\ and\ \citenamefont
  {Greengard}(2013)}]{gimbutas_t-matrix}%
  \BibitemOpen
  \bibfield  {author} {\bibinfo {author} {\bibfnamefont {Z.}~\bibnamefont
  {Gimbutas}}\ and\ \bibinfo {author} {\bibfnamefont {L.}~\bibnamefont
  {Greengard}},\ }\href@noop {} {\bibfield  {journal} {\bibinfo  {journal}
  {Journal of Computational Physics}\ }\textbf {\bibinfo {volume} {232}},\
  \bibinfo {pages} {22} (\bibinfo {year} {2013})}\BibitemShut {NoStop}%
\bibitem [{\citenamefont {Stein}(1961)}]{Stein_1961}%
  \BibitemOpen
  \bibfield  {author} {\bibinfo {author} {\bibfnamefont {S.}~\bibnamefont
  {Stein}},\ }\href@noop {} {\bibfield  {journal} {\bibinfo  {journal}
  {Quarterly of Applied Mathematics}\ }\textbf {\bibinfo {volume} {19}}
  (\bibinfo {year} {1961})}\BibitemShut {NoStop}%
\bibitem [{\citenamefont {Park}\ and\ \citenamefont {Lu}(2007)}]{park_align}%
  \BibitemOpen
  \bibfield  {author} {\bibinfo {author} {\bibfnamefont {J.}~\bibnamefont
  {Park}}\ and\ \bibinfo {author} {\bibfnamefont {W.}~\bibnamefont {Lu}},\
  }\href@noop {} {\bibfield  {journal} {\bibinfo  {journal} {Applied Physics
  Letters}\ }\textbf {\bibinfo {volume} {91}},\ \bibinfo {pages} {053113}
  (\bibinfo {year} {2007})}\BibitemShut {NoStop}%
\bibitem [{\citenamefont {M{\"u}hlig}\ \emph
  {et~al.}(2011{\natexlab{b}})\citenamefont {M{\"u}hlig}, \citenamefont
  {Menzel}, \citenamefont {Rockstuhl},\ and\ \citenamefont
  {Lederer}}]{muhlig_multipole}%
  \BibitemOpen
  \bibfield  {author} {\bibinfo {author} {\bibfnamefont {S.}~\bibnamefont
  {M{\"u}hlig}}, \bibinfo {author} {\bibfnamefont {C.}~\bibnamefont {Menzel}},
  \bibinfo {author} {\bibfnamefont {C.}~\bibnamefont {Rockstuhl}}, \ and\
  \bibinfo {author} {\bibfnamefont {F.}~\bibnamefont {Lederer}},\ }\href@noop
  {} {\bibfield  {journal} {\bibinfo  {journal} {Metamaterials}\ }\textbf
  {\bibinfo {volume} {5}},\ \bibinfo {pages} {64} (\bibinfo {year}
  {2011}{\natexlab{b}})}\BibitemShut {NoStop}%
\bibitem [{\citenamefont {Miller}\ \emph {et~al.}(2013)\citenamefont {Miller},
  \citenamefont {Qiu}, \citenamefont {Joannopoulos},\ and\ \citenamefont
  {Johnson}}]{Miller_comment}%
  \BibitemOpen
  \bibfield  {author} {\bibinfo {author} {\bibfnamefont {O.~D.}\ \bibnamefont
  {Miller}}, \bibinfo {author} {\bibfnamefont {W.}~\bibnamefont {Qiu}},
  \bibinfo {author} {\bibfnamefont {J.~D.}\ \bibnamefont {Joannopoulos}}, \
  and\ \bibinfo {author} {\bibfnamefont {S.~G.}\ \bibnamefont {Johnson}},\
  }\href@noop {} {\bibfield  {journal} {\bibinfo  {journal} {arXiv preprint
  arXiv:1310.1503}\ } (\bibinfo {year} {2013})}\BibitemShut {NoStop}%
\end{thebibliography}

%

\end{document}